\documentclass[fleqn,usenatbib]{mnras} 

\usepackage[T1]{fontenc}
\usepackage{graphicx}	
\usepackage{amsmath,amsfonts,amssymb}
\usepackage{makeidx}
\usepackage{epstopdf} 
\usepackage{booktabs,caption,fixltx2e}
\usepackage[flushleft]{threeparttable}
\usepackage{hyperref}
\usepackage{multirow}
\usepackage{rotating}
\usepackage{color}
\usepackage{soul}
\usepackage{ulem}
\usepackage{ae,aecompl}

\newcommand{\LCDM}{$\Lambda$CDM}

\newcommand{\mtrx}[1]{\textbf{\textsf{#1}}}

\title[Quantifying Tensions between CMB and Distance Datasets]{Quantifying Tensions between CMB and Distance Datasets in Models with Free Curvature or Lensing Amplitude}

\author[S.~Grandis et al.]{
S.~Grandis$^{1,2}$,
D.~Rapetti$^{1,2,3,4}$,
A.~Saro$^{1,2}$,
J.~J.~Mohr$^{1,2,5}$,
J. P. Dietrich$^{1,2}$
\\
$^{1}$ Faculty of Physics, Ludwig-Maximilians-Universit\"at, Scheinerstr.\ 1, 81679 Munich, Germany \\
$^{2}$ Excellence Cluster Universe, Boltzmannstr.\ 2, 85748 Garching, Germany \\
$^{3}$ Center for Astrophysics and Space Astronomy, Department of Astrophysical and Planetary Science, University of Colorado, \\ 
Boulder, C0 80309, USA \\
$^{4}$ NASA Ames Research Center, Moffett Field, CA 94035, USA \\
$^{5}$ Max Planck Institute for Extraterrestrial Physics, Giessenbachstr.\ 85748 Garching, Germany
}

\date{Accepted XXX. Received YYY; in original form ZZZ}

\pubyear{2016}

\begin{document}
\maketitle
\flushbottom

\begin{abstract}
Recent measurements of the Cosmic Microwave Background (CMB) by the Planck Collaboration have produced arguably the most powerful observational evidence in support of the standard model of cosmology, i.e. the spatially flat \LCDM\ paradigm.  In this work, we perform model selection tests to examine whether the base CMB temperature and large scale polarization anisotropy data from Planck 2015 (P15) prefer any of eight commonly used one-parameter model extensions with respect to flat \LCDM.  We find a clear preference for models with free curvature, $\Omega_\mathrm{K}$, or free amplitude of the CMB lensing potential, $A_\mathrm{L}$. We also further develop statistical tools to measure tension between datasets. We use a Gaussianization scheme to compute tensions directly from the posterior samples using an entropy-based method, the surprise, as well as a calibrated evidence ratio presented here for the first time. We then proceed to investigate the consistency between the base P15~CMB data and six other CMB and distance datasets. In flat \LCDM\ we find a $4.8\sigma$ tension between the base P15~CMB data and a distance ladder measurement, whereas the former are consistent with the other datasets. In the curved \LCDM\ model we find significant tensions in most of the cases, arising from the well-known low power of the low-$\ell$ multipoles of the CMB data. In the flat \LCDM$+A_\mathrm{L}$ model, however, all datasets are consistent with the base P15~CMB observations except for the CMB lensing measurement, which remains in significant tension. This tension is driven by the increased power of the CMB lensing potential derived from the base P15~CMB constraints in both models, pointing at either potentially unresolved systematic effects or the need for new physics beyond the standard flat \LCDM\ model.
\end{abstract}


\section{Introduction}

Over the last two decades, growing observational evidence has been collected in support of a model with a flat geometry, cold dark matter (CDM) and a cosmological constant, $\Lambda$. This model has been extremely successful in the face of observational constraints from a wide variety of datasets, such as temperature and anisotropy measurements of the Cosmic Microwave Background \citep[CMB;][]{wmap1, wmap2, planck13, planck15_data_prod, planck15, planck_likelis}. It also accurately predicts measurements of the cosmic distance ladder \citep{riess, aubourg, efstathiou, riess16}, supernovae type Ia \citep{conley, betoule}, baryon acoustic oscillations \citep[BAO;][]{beutler, ross, anderson, delubac}, cluster gas mass fraction \citep{Allen08, Mantz14}, cosmic shear correlation function~\citep{kilbinger, mandelbaum, DES_SV_shear}, CMB lensing~\citep{act_lensing, spt_lensing, planck_lensing}, and cluster number counts \citep{wtg2, Vikhlinin09, Benson13, act_clusters, bocquet1, planck15_clusters, Mantzetal15, TdH16}.

Despite some recently discussed tensions concerning the value of the present day Hubble parameter \citep[see for instance][]{h0_tension_1, h0_tension_2, H0concordance, riess16} and the power of scalar fluctuations as measured from the CMB and large scale structure probes \citep[see, among others,][]{lss_tension_2, lss_tension_1, raveri, SG, dic_WL}, the constraints from all these observations seem to agree reasonably well with each other in this model. 

To test various key assumptions of the flat \LCDM\ model, we consider a series of one parameter extensions to this model and investigate whether the increased complexity of the extended models is needed to improve the goodness of fit to the data. In this work we show that the temperature and large scale polarization CMB anisotropy measurements, i.e. the base CMB constraints, of the Planck Collaboration \citep{planck15, planck_likelis} (hereafter called base P15~CMB) prefer a \LCDM\ model with free curvature, $\Omega_\mathrm{K}$, or free lensing potential amplitude, $A_\mathrm{L}$. Besides model selection, tensions between different datasets within the same model can also indicate that the assumed model is not adequate. We examine the level of agreement between the base CMB constraints and different additional datasets in the flat \LCDM, curved \LCDM, and flat \LCDM$+A_\mathrm{L}$ models. In flat \LCDM, we find mostly consistency among the datasets we consider with the exception of a significant tension with a recent distance ladder measurement, but this is no longer true in the two extended models we examine. We find that the base P15~CMB constraints are in significant tension with most external datasets in curved \LCDM, whereas in flat \LCDM $+A_\mathrm{L}$, only the CMB lensing data show a significant disagreement with the P15~CMB constraints. 

To perform these data consistency tests, we have developed different statistical tools. Generalizing the Gaussianization scheme of \citet{schuhmann}, for each model we consider, we find a transformation of parameter space that maps onto Gaussian distributions both the P15~CMB constraints alone and these combined with one external dataset. We then measure the degree of tension introduced by these combinations using the entropy based `surprise', which was introduced by \citet{seehars1, seehars2} to measure the consistency of an historical sequence of CMB surveys, and employed by \citet{SG} to demonstrate the agreement of different external datasets with the Wilkinson Microwave Anisotropy Probe \citep[WMAP,][]{wmap1, wmap2}. The Gaussianization procedure is crucial to test the consistency between datasets in models with strong parameter degeneracies, as it allows one to analytically approximate their constraints. This provides an important test, which we argue should be systematically performed when combining datasets.

We also investigate the statistical properties of evidence ratios, a widely used measure of dataset agreement \citep[see][]{evidence_ratio1, evidence_ratio2, evidence_ratio3, evidence_ratio4, evidence_ratio5, raveri}. We demonstrate theoretically and with simple examples that evidence ratios can be highly biased and therefore need to be accurately calibrated. We also compare calibrated evidence ratios to the surprise results, and find that they give very comparable measures of the significance of the tension.
  
We organize the paper as follows. In Section~\ref{sec:stat}, we discuss the statistical tools employed. In Section~\ref{sec:data}, we present the datasets used in our analysis. We then report our results on model selection and dataset consistency in Section~\ref{sec:results}, discussing the impact of systematics and choices of priors in Section~\ref{sec:discussion}, which also contains a discussion of the physical effects responsible for the deviation from flatness or from $A_\mathrm{L}=1$.

\section{Statistical Methods}\label{sec:stat}

Cosmological constraints on a specific model, $M$, derived from astrophysical data, $D$, are usually expressed as a posterior distribution $p(\pmb{\theta}|D, M)$ on the space of cosmological parameters $\pmb{\theta}$. Posterior distributions can be obtained by using the Bayes' Theorem as
\begin{equation}
p(\pmb{\theta}|\,D, M) = \frac{L(D|\,\pmb{\theta},M)}{E(D|\,M)} p(\pmb{\theta})\,,
\end{equation}
where $ p(\pmb{\theta})$ is a prior, $ L(D|\,\pmb{\theta},\,M)$ the likelihood and $E(D|M)$ the evidence.

\subsection{Gaussianization}\label{sec:gauss}
In some models, the posterior distribution displays significant departures from Gaussianity. This complicates both a possible analytic approximation of the posterior as well as the comparison with other posterior distributions. However, as explicitly shown by \citet{schuhmann}, a suit of optimized transformations of the parameters can efficiently map a generic uni-modal distribution onto a Gaussian distribution. This allows one to analytically approximate the distribution, significantly speeding up its evaluation. For details on the precision of this approximation, see Appendix \ref{sec:gaussianization}.
   
Here we generalize the Gaussianization method proposed by \citet{schuhmann} to simultaneously Gaussianize two distributions. Such a joint Gaussianization will allow us to compare the two distributions analytically. For details, see also Appendix~\ref{sec:gaussianization}. In the following, we present the statistical tools we employ to quantify comparisons between datasets (Section~\ref{sec:tensions}) and between models (Section~\ref{sec:modelselec}).
  
\subsection{Quantifying Tension}\label{sec:tensions}

Given the variety of cosmological datasets, it is of great importance to asses their mutual agreement. The absence of this agreement is usually referred to as `tension' between datasets. We first discuss an entropy based method to measure these tensions and then an evidence ratio based one.
   
\subsubsection{Entropy Based Method}\label{sec:surprise}
To quantify the consistency of a dataset $D_1$ with another dataset $D_2$ we can use the Kullback--Leibler divergence, also called relative entropy, introduced by \citet{kullbackleibler},
\begin{equation}
KL[D_2|\,D_1] = \int \text{d}^d \pmb{\theta} \, p(\pmb{\theta}|\,D_1,D_2, M) \ln \Bigg ( \frac{p(\pmb{\theta}|\,D_1,D_2, M)}{p(\pmb{\theta}|\,D_1)} \Bigg)\,,
\end{equation}
where $p(\pmb{\theta}|\,D_1)$ is the posterior distribution of the dataset $D_1$, which we employ as a prior for updating the joint posterior of the two datasets $p(\pmb{\theta}|\,D_1,D_2, M)$.

As discussed elsewhere \citep{seehars1, seehars2, SG}, the relative entropy depends on the datasets $D_{1}$ and $D_{2}$, and as such has an expected value $\langle KL \rangle_{D_2|D_1}$ and a mean fluctuation around this value $\sigma(KL)$, which depends on the expected distribution of the dataset $D_2$ given the prior $p(\pmb{\theta}|\,D_1)$. The difference between the actual relative entropy and the expected relative entropy is defined by \citet{seehars1} as the \textit{surprise} $S = KL[D_2|\,D_1] - \langle KL \rangle_{D_2|D_1}$. If the surprise is negative, $S<0$, the dataset $D_2$ is in better agreement with the prior than expected; if the surprise is positive, the dataset $D_2$ is in worse agreement with the prior than expected. Comparing the surprise $S$ to its expected fluctuation $\sigma(KL)$ allows one to estimate the significance of the underlying tension \citep[see][for more details]{seehars1, seehars2}.
    
The relative entropy is invariant under transformations in parameter space \citep[for proof see appendix B in][]{SG}, and it is analytic if prior and posterior are multivariate Gaussian distributions \citep[see][]{seehars1}. Thus, it can be easily estimated for two generic distributions after a joint Gaussianization. As shown by \citet{seehars1, seehars2}, in this case it will be given by
\begin{equation} \label{eq:S}
S = \frac{1}{2} \pmb{\Delta \mu}^T \mtrx{C}^{-1}_\mathrm{pr} \pmb{\Delta \mu} - \frac{1}{2}  \text{tr}\left(\mathbb{I} - \mtrx{C}_\mathrm{po} \mtrx{C}^{-1}_\mathrm{pr}\right),    
\end{equation}
where $\pmb{\Delta \mu}$ is the difference in means of the transformed distributions, $\mtrx{C}_\mathrm{pr}$  and $\mtrx{C}_\mathrm{po}$ the covariances of the transformed prior and posterior respectively, `tr' stands for trace, and $\mathbb{I}$ is the identity matrix. In this case, the variance of the relative entropy is given by $\sigma^2(KL) =  \text{tr}\big \lbrack  ( \mtrx{C}^{-1}_\mathrm{pr} \mtrx{C}_\mathrm{po} - \mathbb{I} )^2 \big \rbrack  /2$. Thus, given estimates of covariances and means for prior and posterior, these quantities can be easily estimated. Note that all entropy based results are given in units of `bits' by normalising with $\ln 2$. 

As can be seen from equation~(\ref{eq:S}), the surprise measures the shift in the mean values $\pmb{\Delta \mu}$ created by the update of $p(\pmb{\theta}|\,D_1)$ with $D_2$, and asses how significant this shift is by comparing it to the expected fluctuation  $\sigma(KL)$. Consequently, it is well suited to test whether $D_2$ should be added to the constraints of $D_1$.
    
\subsubsection{Calibrated Evidence Ratio}
    
A standard way \citep[see][]{evidence_ratio1, evidence_ratio2, evidence_ratio3, evidence_ratio4, evidence_ratio5, raveri} of assessing the degree of agreement between two datasets $D_1$, $D_2$ is given by the so called \textit{evidence ratio}
\begin{equation} \label{eq:R}
R = \frac{E(D_1, D_2)}{E(D_1)\, E(D_2)}\,, 
\end{equation}
where $E(D_1, D_2)$ is the joint evidence of the two datasets $D_1$ and $D_2$, and $E(D_1)$ and $E(D_2)$ are the evidences of the individual datasets. 
    
This ratio is interpreted using the Jeffreys' scale introduced by \citet{jeffreys}, where $\ln R >0 $ indicates agreement and  $\ln R<0 $ indicates inconsistency. However, as pointed out by \citet{seehars2}, this interpretation does not take into account the statistical behaviour of the evidence ratio. For this sake, in Appendix~\ref{sec:evidence} we compute the expected evidence ratio $\langle \ln R \rangle$ and its variance $\sigma^2(R) = \big \langle (\ln R - \langle \ln R \rangle)^2  \big \rangle$ for the case of data described by a Gaussian likelihood under the assumption of a linear model and flat priors, and define the \textit{calibrated evidence ratio} $\ln R - \langle \ln R \rangle$. The latter allows a more quantitative measurement of tension than the somewhat heuristic Jeffreys' scale, and avoids biasing the results. For other details on our treatment of the evidence ratio see Appendix~\ref{sec:app_b}.

\subsection{Model Selection}\label{sec:modelselec}

%
\begin{table}
\caption{Interpretation of the difference in Deviance Information Criterion, $\Delta$DIC, using the Jeffreys' scale as proposed by \citet{dic_1}. For nested models described by uncorrelated Gaussian likelihoods, $\Delta$DIC can be straightforwardly related to the deviation of the additional parameter in the more complex model w.r.t. its fixed value in the simpler one. As a reference, for a given $\Delta$DIC we calculate the offset of an additional parameter measured in standard deviations $\sigma$, and in the corresponding p-value.}   
\vspace{5pt}
\setlength\tabcolsep{5pt} 
\centering           
\begin{tabular}{c c c c}        
\hline\hline                 
$\Delta$DIC & preference & $\sigma$ & p-value \\
\hline
$(-2, 0)$ &  insignificant & $(1.41, 2.00)$ & $(2.28\mathrm{e}{-2}, 7.93\mathrm{e}{-2})$ \\
$(-5, -2)$ & positive & $(2.00, 2.65)$ & $(4.02\mathrm{e}{-3}, 2.28\mathrm{e}{-2})$\\
$(-10, -5)$ & strong & $(2.65, 3.46)$ & $(2.70\mathrm{e}{-4}, 4.02\mathrm{e}{-3})$ \\
$(-\infty, -10)$ & decisive & $(3.46, \infty) $ & $(0, 2.70\mathrm{e}{-4})$\\
\hline

\end{tabular}
\label{table:jeffreys}
\end{table}
%
   
To determine whether a given dataset, $D$, prefers a model $M_1$ or model $M_2$, we rely on the \textit{Deviance Information Criterion} (hereafter DIC). Considering the generalized chi-squared $\chi^2(\pmb{\theta}) = -2 \ln L(D|\,\pmb{\theta},M_i)$, the mean goodness of fit over the posterior volume can be estimated as $\big \langle\chi^2\big \rangle = -2 \big \langle\ln L(D|\,\pmb{\theta},M_i)\big \rangle$. A model which fits the data better will have a lower $\big \langle\chi^2\big \rangle$. Motivated by information theory, \citet{dic_1} define the DIC as
\begin{equation}\label{eq:dic}
\text{DIC}(M_i) =  \big \langle\chi^2\big \rangle + p_D\,.
\end{equation}
This balances the mean goodness of fit $ \big \langle\chi^2\big \rangle$ with the Bayesian complexity $p_D$, which measures the effective complexity of the model and is given by
\begin{equation}
p_D = \big \langle\chi^2\big \rangle - \chi^2(\pmb{\tilde \theta})\,,
\end{equation}
where $\pmb{\tilde \theta}$ denotes the maximum likelihood point. A lower DIC means either that the model fits the data better (lower $ \big \langle\chi^2\big \rangle$) or that it has a lower level of complexity, $p_D$. A higher complexity, such as additional model parameters, can only be compensated if they allow a sufficient improvement of the goodness of fit.
        
For model selection, the difference $\Delta \text{DIC} = \text{DIC}(M_2) - \text{DIC}(M_1)$ is interpreted using the Jeffreys' scale (see Table~\ref{table:jeffreys}). $\Delta \text{DIC} =0$ means that the data provide no preference for one model over the other, $-2<\Delta \text{DIC}<0$ that there is `no significant' preference for $M_2$, $-5<\Delta \text{DIC}<-2$ a `positive' preference, $-10<\Delta \text{DIC}<-5$ `strong', and $-10<\Delta\text{DIC}$ `decisive'. The same values but positive indicate a preference for $M_1$ instead.

As an example of model selection with the DIC, consider the following case. Let the data be described by a standardized Gaussian likelihood $-2\ln L = \sum_{i=1}^n \theta_i^2$, where $\theta_1, ..., \theta_n$ are the model parameters of $M_2$, and let the simpler model $M_1$ derive from $M_2$ by setting one of the model parameters $\theta_j$ to the value $\sigma$. In this case, assuming flat priors, the $\Delta\text{DIC}$ can be calculated analytically as $2-\sigma^2$ and the p-value of the offset $\sigma$ can be computed from the fact that the posterior of $\theta_j$ given the data $D$, $p(\theta_j|D,\,M_2)$, is Gaussian. For reference, we present these results in Table~\ref{table:jeffreys}.
    
As discussed by  \citet{dic_1}, the DIC can also deal with strong parameter degeneracies, such as the geometrical degeneracy of the CMB data in curved models. It takes also into account `parameter volume effects', as it considers the goodness of fit averaged over the posterior volume. Furthermore, this measure can be easily computed from a posterior sample, which saves the values $\ln L(D|\,\pmb{\theta},M_i)$ in every point, making it more versatile than the evidence ratio \citep[for applications of this measure to astrophysics and cosmology, see][]{dic_quasars, dic_wmap, dic_clusters, dic_WL}.

\section{Cosmological Data} \label{sec:data}

\subsection{Planck Data}
We employ the \texttt{TT\_lowTEB} constraints from the Planck Collaboration \citep{planck15} of the temperature and large scale polarization anisotropies in the CMB, which we also refer to as `base P15~CMB'.  When considering the full Planck 2015  temperature and polarization measurements, we use the \texttt{TTTEEE\_lowTEB} sample, which we will also refer to as `full P15~CMB'. We also use the CMB lensing constraints \citep{planck_lensing} included in the \texttt{TT\_lowTEB+lensing} samples, referring to them as `CMB lens'. The Monte Carlo Markov Chain (MCMC) CMB samples analyzed in this work were downloaded from the Planck Legacy Archive\footnote{\url{http://pla.esac.esa.int/pla/\#cosmology}} and subsequently Gaussianized as described in Section~\ref{sec:gauss}.

\subsection{Additional Geometrical Probes}\label{sec:add}
Given an analytic expression for the base P15~CMB constraints derived from the Gaussianization process described in Section~\ref{sec:gauss} and Appendix~\ref{sec:gaussianization}, we can easily combine them with measurements from geometrical probes. This has the advantage that the prominent geometrical degeneracy of the CMB data in curved models can be broken \citep[see e.g.][]{geom_deg_1, geom_deg_2}. We 
compute the theoretical distance predictions using \texttt{CAMB} \citep{Lewis:2008wx}\footnote{\url{http://camb.info/}} and sample the joint constraints with the parallelized MCMC engine \texttt{CosmoHammer} \citep{cosmohammer}\footnote{\url{https://github.com/cosmo-ethz/CosmoHammer}}. In the following we present the additional geometrical datasets we used in this work. 

\subsubsection{Datasets}
Various recent constraints on the Hubble constant $H_0$ exist in the literature \citep{riess, H0concordance, aubourg, efstathiou, riess16}. In the present work, we use the latest result by \citet[][hereafter R16]{riess16}, who obtain $H_0$ $= 73.02 \pm 1.79\, \text{km}\,\text{s}^{-1}\,\text{Mpc}^{-1}$. As a consistency check, we also use the constraint $H_0^\text{E14}= 70.6 \pm 3.3\, \text{km}\,\text{s}^{-1}\,\text{Mpc}^{-1}$ reported by \citet[][hereafter E14]{efstathiou}. We use these measurements as Gaussian likelihoods. This simple form will also allow us to employ them to compute evidences as described in Appendices~\ref{sec:evidence} and~\ref{sec:est_ev}.

We also use measurements of the Hubble parameter as a function of redshift from the latest calibration of a large compilation of supernovae type Ia (SNe) data by \cite{betoule}. This work combines observations from the Supernovae Legacy Survey, the Sloan Digital Sky Survey (SDSS) and the Hubble Space Telescope, and provides a binned version of the SNe Hubble diagram with the corresponding covariance matrix. As shown in appendix E of \citet{betoule}, computing the luminosity distance in $\text{Mpc}\,h^{-1}$, marginalising analytically over the intrinsic luminosity of the SNe and assuming a Gaussian likelihood allows a straightforward computation of the SNe constraints. 

We also include constraints from baryon acoustic oscillations (BAO) derived from galaxy correlations in the 6dF Galaxy Survey by \citet{beutler}, the SDSS main galaxy sample by \citet{ross}, and the  Baryon Oscillation Spectroscopic Survey (BOSS) by \citet{anderson}. The Planck Collaboration \citep[][see e.g. page 24]{planck15} provided samples of these BAO measurements together with the base CMB data, labelled as \texttt{TT\_lowTEB+BAO}.
       
\citet{delubac} derived BAO measurements from the Ly$\alpha$ forest in the Data Release 11 of BOSS. We will refer to this measurement as `Ly$\alpha$ BAO'. These results are reported as $D_\text{A}(z=2.34) = 1662\pm 96\, \text{Mpc}\, (r_\text{d}/r_\text{fid})$ and $H(z=2.34) = 222\pm 7\, \text{km}\,\text{s}^{-1}\,\text{Mpc}^{-1}\, (r_\text{fid}/r_\text{d})$, where $D_A$ is the angular diameter distance, $H(z)$ the expansion rate at a given redshift $z$, $r_\text{fid} = 147.4 \, \text{Mpc}$ the fiducial sound horizon used by \citet{delubac} and $r_\text{d}$ the sound horizon dependent on the cosmological parameters. We assume Gaussian likelihoods for these results.


\begin{figure}

\hskip-0.15in
\includegraphics[width=0.54\textwidth]{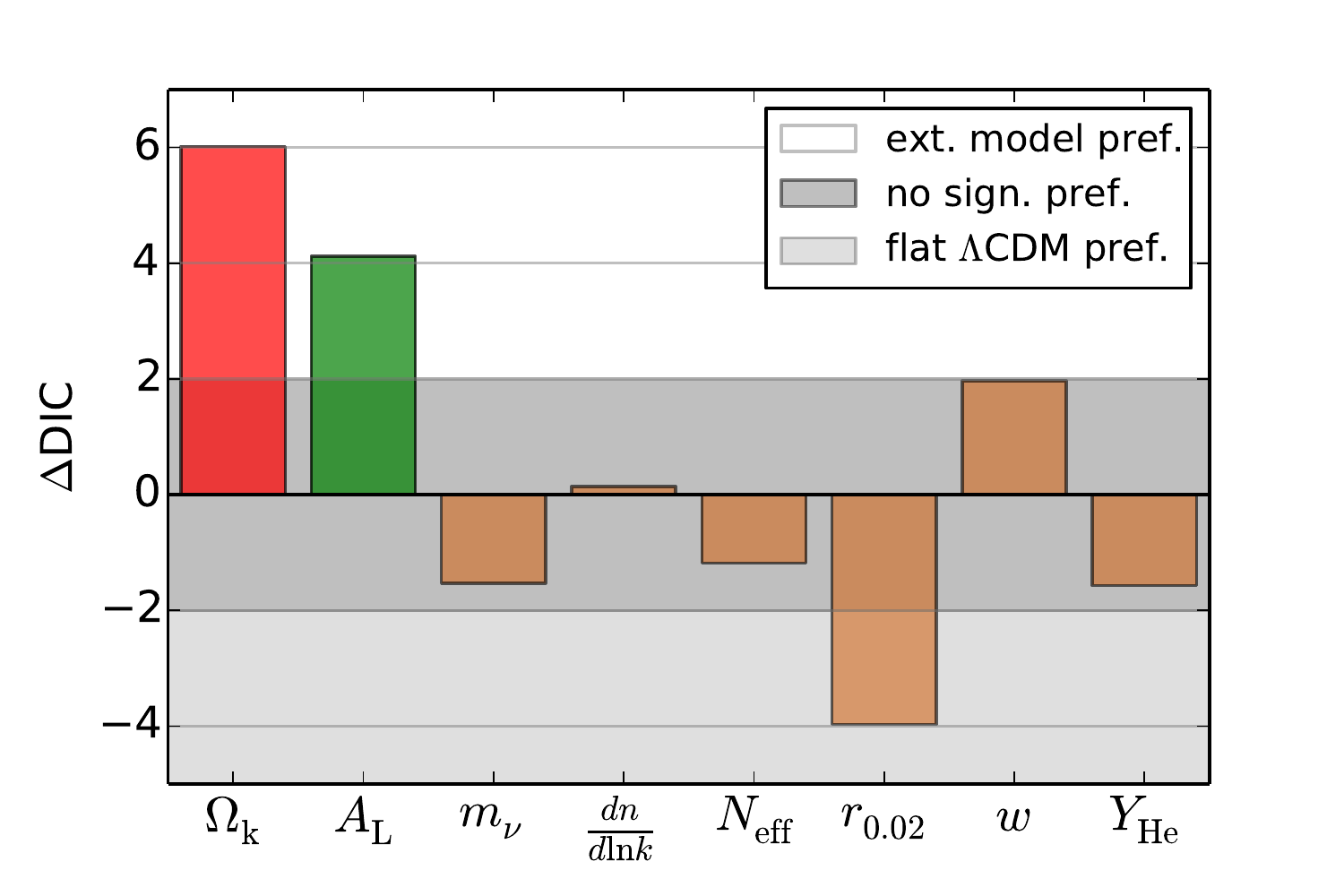}
\vskip-0.in
\caption{Differences in Deviance Information Criterion, $\Delta \text{DIC}$, between flat \LCDM\ and various one-parameter extensions of this model.  These results are estimated from the publicly available \texttt{TT\_lowTEB} constraints.  The ranges $-2<\Delta \text{DIC}<2$, $\Delta \text{DIC}>2$ and $\Delta \text{DIC}<-2$  indicate no significant preference for either model, a preference for the extended model, or that the data prefer the simpler model, respectively.  Remarkably, we find clear preferences for two of the extended models, flat \LCDM$+A_\mathrm{L}$ (green) and curved \LCDM\ (red).}

\label{fig:dics}
\end{figure}

%

\section{Results}\label{sec:results}

\subsection{Which model is preferred by the P15~CMB data?}\label{sec:model_select}
    
We compute the change in the Deviance Information Criterion $\Delta$DIC between the standard flat \LCDM\ and several extended models. We consider all the one-parameter extensions for which the Planck Collaboration published \texttt{TT\_lowTEB} constraints, namely: $\Omega_\mathrm{K}$ (we refer to this model as \textit{curved} \LCDM); $A_\mathrm{L}$, the amplitude of the CMB lensing potential (we refer to this model as flat \LCDM$+A_\mathrm{L}$); $m_\nu$, the effective sum of neutrino masses; $\mathrm{d}n/\mathrm{d}\ln k$, the running of the spectral index of scalar perturbations; $N_\mathrm{eff}$, the effective number of relativistic degrees of freedom; $r_{0.02}$, the tensor to scalar mode ratio; $w$, the dark energy equation of state parameter; and $Y_\mathrm{He}$, the primordial Helium fraction.
    
In Fig.~\ref{fig:dics} and Table~\ref{table:dics}, we show the differences between the DIC of flat \LCDM\ and those of the extended models as calculated from the publicly available samples. We find that the P15~CMB data favor most the curved \LCDM\ model ($\Delta \text{DIC} = 6.02$), followed by the model with free $A_\mathrm{L}$ ($\Delta \text{DIC} = 4.12$). For the other model extensions we find no significant preference over flat \LCDM. We also find that flat \LCDM\ is preferred over a model with free tensor to scalar ratio, $r_{0.02}$.

%
\begin{table}
\caption{$\Delta \text{DIC}$ between flat \LCDM\ and various one-parameter extensions of this model, for the base P15~CMB constraints. The ranges $-2<\Delta \text{DIC}<2$, $\Delta \text{DIC}>2$ and $\Delta \text{DIC}<-2$  indicate no significant preference for either model, a preference for the extended model, or that the data prefer the simpler model, respectively.}   
\vspace{5pt}
\setlength\tabcolsep{5pt} 
\centering           
\begin{tabular}{l c c c c c c c}        
\hline\hline                 
$\Omega_\mathrm{K}$ & $A_\mathrm{L}$ & $m_\nu$ & $\text{d}n/\text{d}\ln k$ &$N_\mathrm{eff}$ & $r_{0.02}$ & $w$ & $Y_\mathrm{He}$ \\     
\hline                     
$6.02$ & $4.12$ & $-1.53$ & $0.14$ & $-1.18$ & $-3.97$ & $1.97$ & $-1.57$ \\      
\hline                        
\end{tabular}
\label{table:dics}
\end{table}
%

The clear preferences for curved \LCDM\ and flat \LCDM$+A_\mathrm{L}$ are related to the fact that both $\Omega_\mathrm{K}$ and $A_\mathrm{L}$ deviate more than 2$\sigma$ from their assumed value in flat \LCDM\ \citep[see also discussion on pgs. 24 and 38 of][]{planck15}. For the case of curved \LCDM, we find a preference for a closed Universe ($\Omega_\mathrm{K}<0$), with a $p$-value of 
\begin{equation}\label{eq:conf}
P(\Omega_\mathrm{K}\geq0)=\int^{\infty}_{0} p(\Omega_\mathrm{K}|\,\texttt{TT\_lowTEB} )\, \text{d}\Omega_\mathrm{K} = 0.0033,
\end{equation}
corresponding to a 2.7$\sigma$ significance. For the flat \LCDM$+A_\mathrm{L}$ model, we find a preference for $A_\mathrm{L}$ larger than $1$, with $p$-value
\begin{equation}\label{eq:conf_alens}
P(A_\mathrm{L} \leq 1) = \int_{-\infty}^{1} p(A_\mathrm{L}|\,\texttt{TT\_lowTEB} )\, \text{d}A_\mathrm{L} = 0.0098,
\end{equation}
which corresponds to a $2.3 \sigma$ deviation from the theoretically expected value $A_\mathrm{L}=1$. 

Even though the P15 CMB likelihood is more complex than the one we use to calculate the reference results presented in Table~\ref{table:jeffreys}, the significances of the offsets and the related p-values we obtain for the curved \LCDM\ and flat \LCDM$+A_\mathrm{L}$ models are consistent with the corresponding $\Delta$DIC values in Table~\ref{table:dics}. We find no significant detection of curvature or $A_\mathrm{L}>1$, although our $\Delta$DIC results indicate significant improvements of the fits w.r.t. the flat \LCDM\ model, which according to the Jeffreys' scale dominate over the increased complexity of the curved and flat \LCDM$+A_\mathrm{L}$ models. This is confirmed by \citet{planck15}, which finds that the mean chi-squared $\langle \chi^2 \rangle$ of the P15 CMB fit for these two models is lower than for flat \LCDM. As discussed in detail in \citet[][pgs. 24 and 38]{planck15}, these two models are sensitive to the large angular scale part of the \texttt{TT} P15~CMB spectrum and the power of CMB lensing potential $C_\ell^{\phi\phi}$, as we show in Section~\ref{sec:physics}. 

\subsection{Quickly Resampling the Planck Constraints}\label{sec:resamp}
The Gaussianization procedure effectively provides an analytic approximation to the P15~CMB likelihood. As we only Gaussianize the constraint on the cosmological parameters, we reconstruct the P15~CMB likelihood marginalized over the nuisance parameters. This is especially useful when using the P15~CMB constraints as priors to be combined with other probes, because it avoids the resampling of the P15 nuisance parameters, significantly reducing the number of parameters involved in this calculation. For example, for flat \LCDM, the \texttt{TT+lowTEB} likelihood depends on 21 parameters, whereas only 5 are the cosmological parameters we resample. These cosmological parameters are $H_0$, the present-day physical baryon and cold dark matter densities in units of the critical density, $\Omega_\mathrm{b} h^2$ and $\Omega_\mathrm{cdm} h^2$, where $h=H_0/100$, and the amplitude and spectral index of the primordial scalar fluctuations, $\ln(10^{10}\, A_\mathrm{s})$ and $n_\mathrm{s}$\footnote{For simplicity, when combining with other datasets, we consider $H_0$ instead of $\theta_\mathrm{MC}$, the ratio of the approximate sound horizon to the angular diameter distance at recombination. The impact of this choice is discussed in Section~\ref{sec:prior}.}.  The remaining 16 parameters include the optical depth to reionization, $\tau$, and 15 nuisance parameters. 

Furthermore, a single call to the analytic likelihood approximation takes less than a milli-second, compared to several seconds for the original Planck likelihood. This opens the possibility to quickly resample the P15~CMB constraints and to efficiently combine them with other probes. For further details, see Appendix~\ref{sec:gaussianization}. The likelihoods are available at the following URL: \url{https://bitbucket.org/grandiss45/gaussianization/}.

\begin{figure*}

\includegraphics[width=0.49\textwidth]{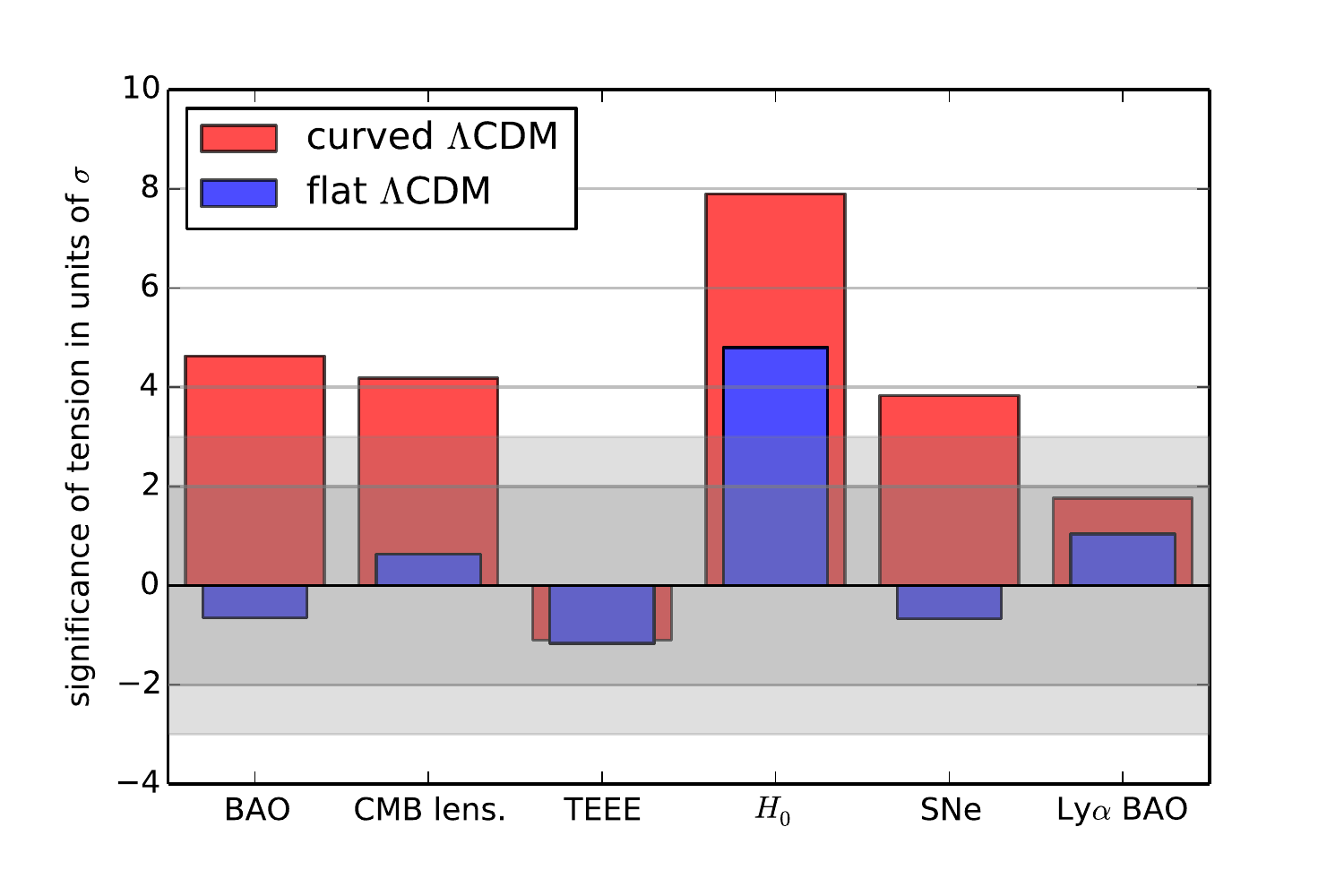}
\includegraphics[width=0.49\textwidth]{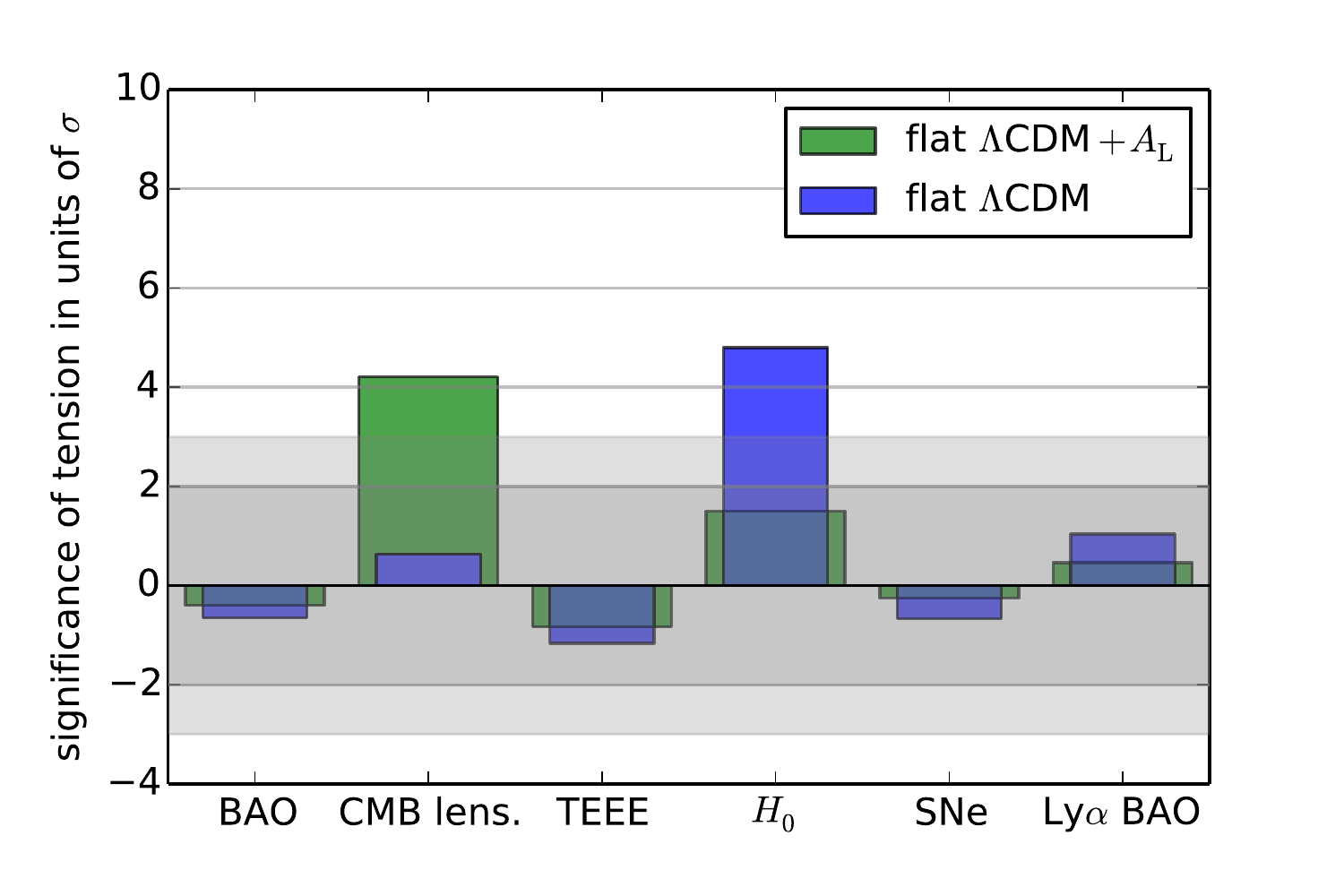}

\caption{Significances of the surprises in units of $\sigma$ in the flat \LCDM\ (blue), curved \LCDM\ (red, left panel), and flat \LCDM$+A_\mathrm{L}$ (green, right panel) models when combining the base P15~CMB constraints with six other probes. The grey regions show the 2$\sigma$ and 3$\sigma$ regions. Surprises more significant than 3$\sigma$ (above the grey regions) indicate tensions of the additional data with the CMB prior. We see that in flat \LCDM\ all probes are consistent with the base P15~CMB constraints, except for the distance ladder measurements. In curved \LCDM, BAO, CMB lensing, $\rm H_0$ and SNe are in significant tension with the base P15~CMB constraints. In flat \LCDM$+A_\mathrm{L}$, CMB lensing is in significant tension with the base P15~CMB constraints, whereas the other probes are in agreement.}
\label{fig:surprises}
\end{figure*}
%

The Gaussianization of the samples is not only helpful to approximate and quickly resample the P15~CMB constraints. It is crucial to computing the surprise analytically. This is possible because the relative entropy is invariant under parameter transformation and is analytic for Gaussian constraints. This allows us to compute the expected relative entropy $\langle KL \rangle_{D_2|D_1}$ and a mean fluctuation around this value $\sigma(KL)$ analytically. As these quantities are obtained by averaging over the distribution of data $E(D_2|D_1)$, it would be very difficult to compute them numerically. The same holds true for the calibration of the evidence ratio $\langle \ln R \rangle$. These integrals over the data are analytic if the constraints can be assumed to be Gaussian, as shown explicitly in Appendix~\ref{sec:evidence}.

\subsection{Adding External Data to the Planck CMB}
Here we test the consistency between each of the datasets described in Section~\ref{sec:data} and the base P15~CMB constraints, first for the standard flat \LCDM\ model and then for the two models that we found in Section~\ref{sec:model_select} to be favored by the base P15~CMB data, i.e. curved \LCDM\ and flat \LCDM$+A_\mathrm{L}$. For the former case, we use the standard set of cosmological parameters listed in Section~\ref{sec:resamp}, while marginalizing over the other parameters sampled by P15 as they are unconstrained by the additional data. In the curved model we also consider the constraints on $\Omega_\mathrm{K}$, whereas in the flat \LCDM$+A_\mathrm{L}$ model we add the parameter $A_\mathrm{L}$.

%
\begin{table}
\caption{Surprise values $S$, expected fluctuations $\sigma$, and significances of tensions $S$/$\sigma$ for different datasets added to the P15 \texttt{TT\_lowTEB} constraints in the models we considered.}  
\vspace{5pt}            
\label{table:surprise}     
\centering                        
\begin{tabular}{c r r r r r r}    

\hline \hline
& BAO & CMB len. & TEEE & $H_0$ & SNe & Ly$\alpha$ BAO \\  
\hline\hline
\multicolumn{7}{l}{flat \LCDM} \\
                
\hline                    
$S$ & $-0.44$ & $0.45$ & $-1.13$ & $1.11$ & $-0.10$ & $0.05$ \\      
$\sigma$ & $0.68$ & $0.72$ & $0.97$ & $0.23$ & $0.15$ & $0.05$ \\
$S$/$\sigma$ & $-0.65$ & $0.63$ & $-1.16$ & $4.78$ & $-0.67$ & $1.04$ \\
\hline\hline
\multicolumn{7}{l}{curved \LCDM} \\                              
\hline                       
$S$ & $6.33$ & $5.45$ & $-1.19$ & $7.36$ & $2.85$ & $0.77$ \\      
$\sigma$ & $1.37$ & $1.30$ & $1.09$ & $0.94$ & $0.74$ & $0.44$ \\
$S$/$\sigma$ & $4.63$ & $4.18$ & $-1.10$ & $7.87$ & $3.83$ & $1.76$ \\
\hline \hline
\multicolumn{7}{l}{flat \LCDM +$A_\mathrm{L}$} \\
\hline                             
$S$ & $-0.31$ & $3.73$ & $-0.92$ & $0.57$ & $-0.10$ & $0.07$ \\      
$\sigma$ & $0.77$ & $0.89$ & $1.11$ & $0.37$ & $0.40$ & $0.16$ \\
$S$/$\sigma$ & $-0.40$ & $4.21$ & $-0.83$ & $1.52$ & $-0.25$ & $0.46$ \\
\hline\hline                                
\end{tabular}

\end{table}
%

\subsubsection{Flat \LCDM}

In flat \LCDM, the base P15~CMB constraints are very well approximated by a multivariate Gaussian distribution, so no Gaussianization is required for resampling. We approximate the constraints directly as multivariate Gaussians, update them with constraints from external data, and then compute the surprise. We summarize our results in Table~\ref{table:surprise} and show them in Fig.~\ref{fig:surprises} (blue bars). We find that for flat \LCDM\ all external datasets are consistent with the base P15~CMB measurements. However, the $H_0$ measurement of R16 is in almost 5$\sigma$ tension with the base P15~CMB dataset\footnote{R16 report that the distance between their mean $H_0$ value and the mean value obtained from the P15 analysis is $3 \sigma$, where $\sigma^2=\sigma_\mathrm{R16}^2 + \sigma_\mathrm{P15}^2$ and $\sigma_\mathrm{P15, R16}$ are the measurement uncertainties on $H_0$ of the two experiments. This result is not in contradiction with our claim, as we instead compute the significance of such a shift. We find that this $3 \sigma$ shift is significant at almost a $5 \sigma$ level. This is also confirmed by our calibrated evidence ratio calculation below.}. 
Worth mentioning is also the tendency to negative surprises for the BAO and SNe data and most strongly for the `\texttt{TEEE}' polarization data. Negative surprises mean that the additional data agree with the prior more than statistically expected. However, these negative surprises are not significant, and can thus be interpreted as statistical fluctuations.

%
\begin{figure}
\hskip-0.2in
\includegraphics[width=0.5\textwidth]{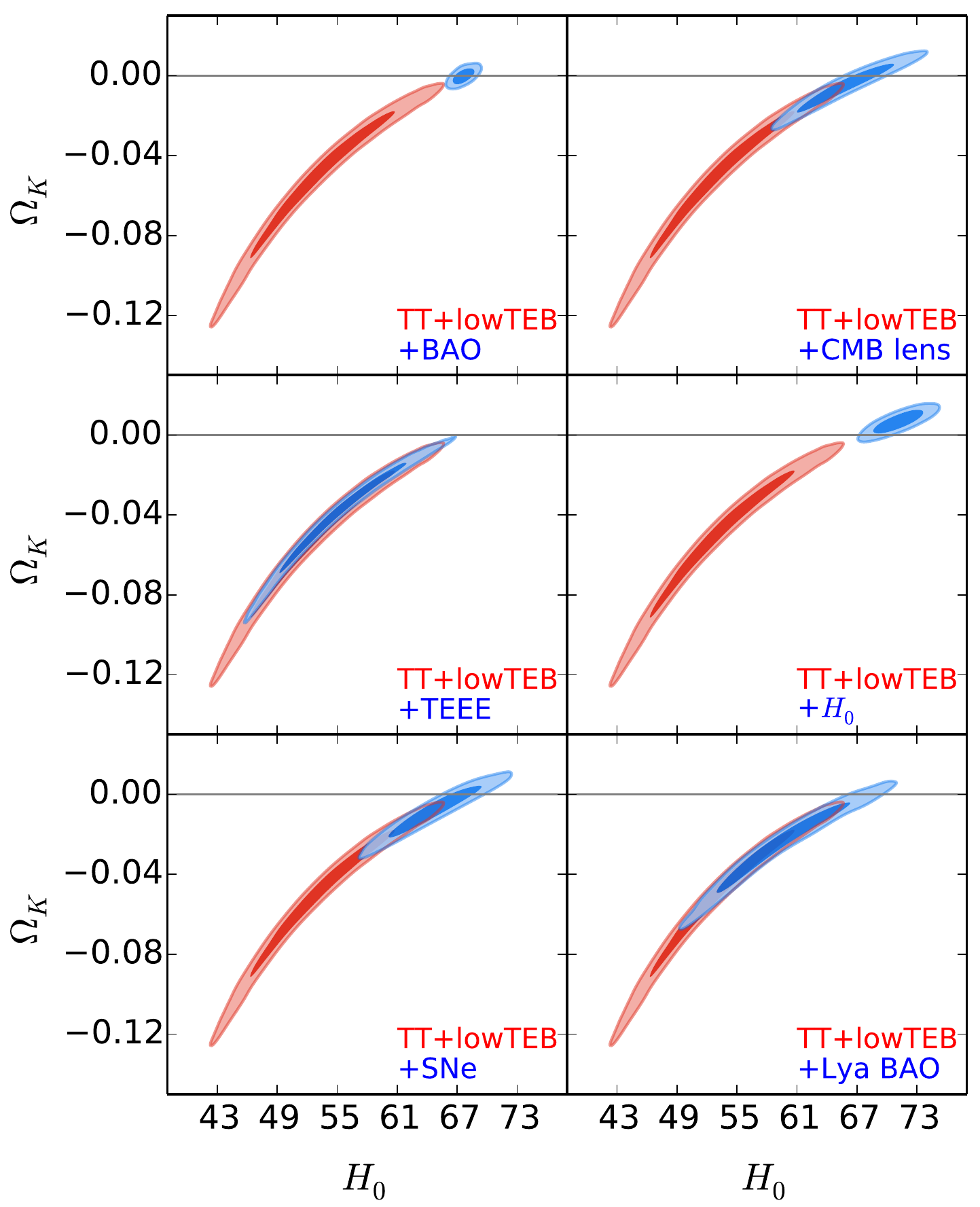}
\vskip-0.in
\caption{Marginal constraints on $H_0$ and $\Omega_\mathrm{K}$ from the base P15~CMB dataset (red contours) and the addition of different datasets to the latter (blue). Adding the P15 small scale polarization data (\texttt{TEEE}) results produces no significant shift of the constraints. However, all external datasets shift the constraints back to flatness, at the cost of increasing tension with the base CMB measurements.}
\label{fig:curv_constr}
\end{figure}

%

\subsubsection{Curved \LCDM}

Fig.~\ref{fig:curv_constr} shows joint constraints in the space of $H_0$ and $\Omega_\mathrm{K}$ for the base P15~CMB dataset alone (red contours) and in separate combinations with six additional datasets (blue contours). As is clear in this figure and as already presented in equation~(\ref{eq:conf}), the base P15~CMB data favor a model with negative curvature at the 2.7$\sigma$ confidence level. In itself, this is not a detection of curvature. Hence, to improve the constraints, additional datasets can be added. Fig.~\ref{fig:curv_constr} shows the impact of such combinations and illustrates how the addition of CMB lensing, two flavors of BAO, SNe and $H_0$ measurements push the P15~CMB constraints noticeably back toward flatness. 
        
By jointly Gaussianizing the prior (the base P15~CMB constraints) and the posterior (combined constraints) for each dataset we add, we transform the cosmological parameters into a space where both distributions are well described by Gaussian distributions. In this space, we estimate the surprise values given in Table~\ref{table:surprise} and shown in Fig.~\ref{fig:surprises} (red bars in the left panel). For an discussion of the accuracy of this method, see Appendix~\ref{sec:gaussianization}. As anticipated by the large shifts in the marginalized plane of $H_0$ and $\Omega_\mathrm{K}$, most additional probes are in significant tension with the base P15~CMB constraints: $H_0$ data at the $8\sigma$ level, BAO and CMB lensing data just over $4\sigma$, and SNe slightly less than $4\sigma$. Ly$\alpha$ BAO data also shift the CMB constraints, but this shift is only a bit less than 2$\sigma$ significant. Finally, also in this model, the \texttt{TEEE} spectrum of the P15 polarization measurements agree with the base P15 constraints more than statistically expected, although not in a statistically significant manner.  

The large surprises in four of the external datasets when combined with the P15~CMB constraints is evidence of significant tensions among the datasets. Thus, our analysis emphasizes that while the combined constraints (P15~CMB + external dataset) prefer flatness more than the P15~CMB dataset alone, this comes at the cost of combining datasets that in four cases are significantly in tension with one another.

\subsubsection{Flat \LCDM$+A_\mathrm{L}$}

Considering the highly significant tensions we find in the curved \LCDM\ model, we also investigate the consistency of the different datasets with the base P15~CMB data in the flat \LCDM$+A_\mathrm{L}$ model.  We show our results in Fig.~\ref{fig:surprises} (green bars in the right panel) and Table~\ref{table:surprise}. Contrary to the curved \LCDM\ model, we find that all distance measures are in good agreement with the base P15~CMB constraints. In the case of the $\rm H_0$ measurement, we find that the significance of tension is reduced from $4.8 \sigma$ in flat \LCDM\ to $1.5 \sigma$ in the flat \LCDM$+A_\mathrm{L}$ model. This is to some extent unsurprising, as these datasets do not directly constrain the additional parameter $A_\mathrm{L}$. But it is worth noting that leaving the $A_\mathrm{L}$ parameter free in the CMB fit, does not change the constraints on the other parameters in a way that is inconsistent  with the various distance measure datasets. Actually, it allows for higher values of $H_0$, reducing the tension with the distance ladder measurements.
    
However, CMB lensing measurements are sensitive to the lensing of the CMB by construction. This dataset shows a tension of $4\sigma$ with the base P15~CMB data. This tension is driven by the constraints on the lensing amplitude. As shown by the Planck Collaboration \cite[pg. 24]{planck15} the constraints from the base CMB ($A_\mathrm{L}=1.22\pm0.10$) are shifted strongly when the CMB lensing data are added ($A_\mathrm{L}=1.04\pm0.06$). The latter is an indication that two datasets which are inconsistent with each other have been combined. We will discuss the underlying physical description of these constraints in Section~\ref{sec:physics}.
    
\subsection{Another Independent Measurement of Tension}
%
\begin{table}
\caption{Evidence ratio results for some of the datasets. $\ln \hat R$ denote the numerical and $\ln R$ the analytic estimates respectively.  $\langle \ln R \rangle$ is the calibration of the evidence ratio and  $\ln R - \langle \ln R \rangle$ the calibrated evidence ratio. `Sig' stands for the significance $(\ln R - \langle \ln R \rangle)/\sigma(\ln R)$, where in one dimension $\sigma(\ln R)=1/\sqrt{2}$. Note that contrary to the surprise values, in the case of evidence ratios negative values indicate tension and positive values indicate agreement.}   
\vspace{5pt}    
\label{table:evidences}     
\centering                      
\begin{tabular}{l r r r r r}
   
\hline\hline                
Flat & $\ln \hat R$ & $\ln R $ & $\langle \ln R \rangle$ & $\ln R - \langle \ln R \rangle$ & Sig \\   
\hline                       
$H_0$ R16 & $-5.6 \pm 1.9$ & $-5.59$ & $-2.13$ & $-3.46$ & $-4.89$ \\ 
$H_0$ E14 & $-2.6 \pm 0.3$ & $-2.61$ & $-2.65$ & $0.04$ & $0.06$ \\
SNe, flat & $2.2 \pm 0.2$ & $2.28$ & $1.89$ & $0.39$ &  $0.54$ \\
 \hline
\hline
Curved & $\ln \hat R$ & $\ln R $ & $\langle \ln R \rangle$ & $\ln R - \langle \ln R \rangle$ & Sig \\ 
\hline
$H_0$ R16 & $-9.2 \pm 3.7$ & $-9.39$ & $-3.01$ & $-6.30$ & $-8.90$ \\
$H_0$ E14 & $-6.6 \pm 1.5$ & $-6.85$ & $-3.22$ & $-3.63$ & $-5.13$ \\
\hline                                 
\end{tabular}
\end{table}
%
As a consistency check for our results, we also employ evidence ratios.  We compute the evidence ratios analytically (see Appendix~\ref{sec:evidence}) for those datasets and models where the likelihood of the data could be assumed to be a simple Gaussian. We use special care in calibrating the analytic evidence ratio $\ln R -\langle \ln R \rangle$, as discussed in Appendix~\ref{sec:evidence}. We also validate our analytic computations with numerical estimates, $\ln \hat R$ (see Appendix~\ref{sec:est_ev}), which allow us to relax the assumption of Gaussianity for the base P15~CMB likelihood. We summarize our findings in Table~\ref{table:evidences}.

\begin{figure*}
\hskip-0.10in
\includegraphics[width=0.8\textwidth]{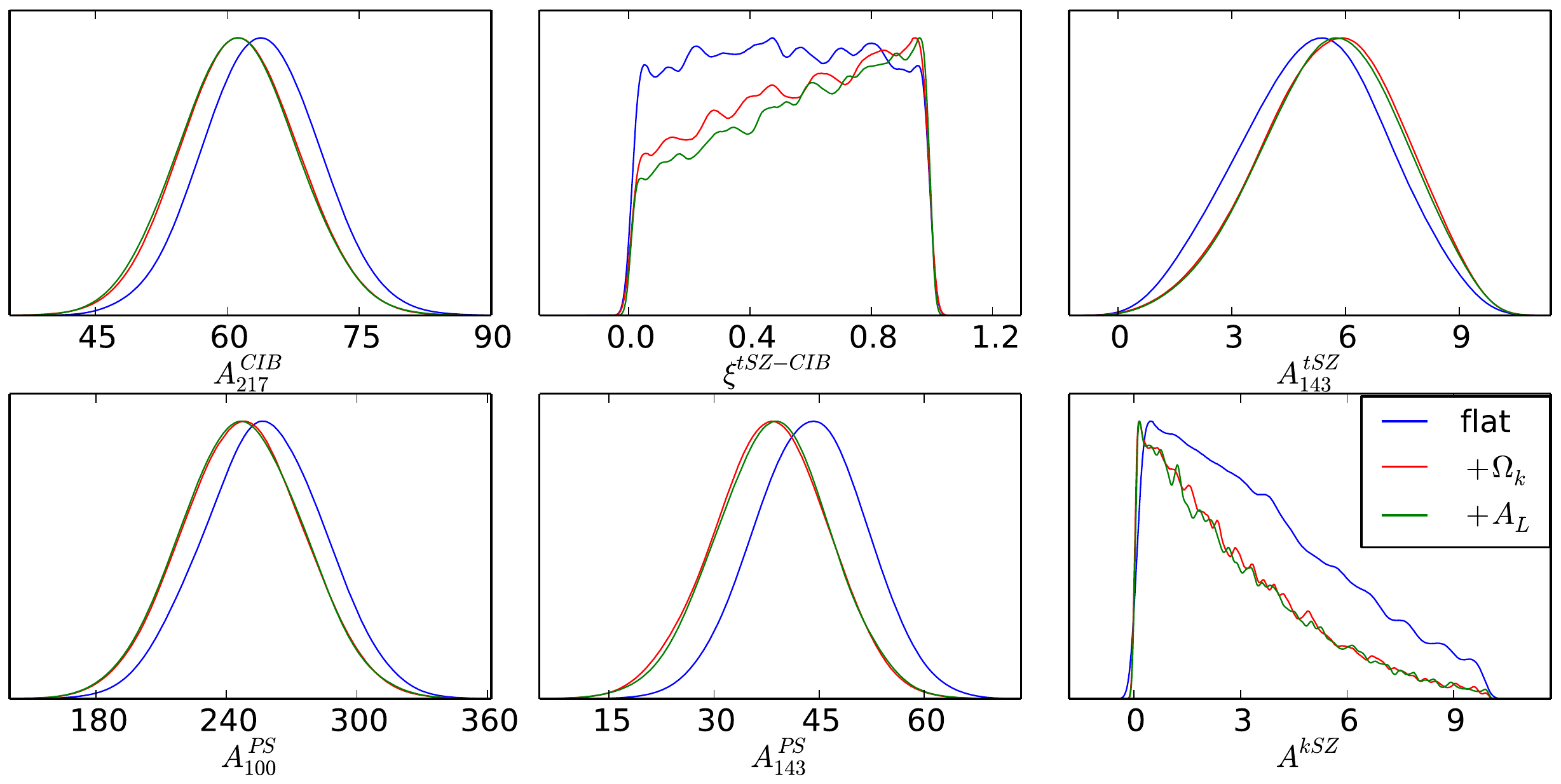}
\vskip-0.10in
\caption{Constraints on nuisance parameters of the base P15~CMB data in flat \LCDM, curved \LCDM, and flat \LCDM$+A_\mathrm{L}$. For simplicity, we include only the marginalized constraints on the parameters that display the largest changes with respect to the flat \LCDM\ model. However, no major shifts are present in these nuisance parameters. Interestingly, both extensions, free $\Omega_\mathrm{K}$ and $A_\mathrm{L}$, shift these constraints in a very similar manner.}
\label{fig:nuisance}
\end{figure*}
%

We find that the numerical evidence $\ln \hat R$ and the analytic evidence $\ln R$ agree. Importantly, in most of the cases we find that the expected evidence ratio $\langle \ln R \rangle$ is very different from zero. Not accounting for the correct calibration can therefore lead to a serious mis-estimation of the degree of tension, as can be seen in the case of $H_0$ E14 and SNe for flat \LCDM. Both agree with the base P15~CMB data, as seen with both the surprise $S$ and the calibrated evidence ratio $\ln R - \langle \ln R \rangle$. However, just considering the evidence ratio $\ln R$ would have biased our conclusion, leading to an overestimation of the agreement in the case of SNe and an underestimation of the agreement in the case of $H_0$ E14. We conclude from this simple example, that uncalibrated evidence ratios can be significantly biased, as discussed further in Appendix~\ref{sec:evidence}.
    
Considering the calibrated evidence ratios $\ln R - \langle \ln R \rangle$ in Table~\ref{table:evidences} we detect the same tensions as with the surprise (see Tables~\ref{table:surprise} and~\ref{table:h0s}). Furthermore, the calibrated evidence ratio, which scatters with $\sigma(\ln R)=1/\sqrt{2}$, have significances comparable to the significances of the surprise. We conclude that in these examples the two measures of tension give very similar results, despite the fact that they detect tensions in different ways, as discussed in Appendix \ref{sec:evidence}. This is reassuring for our primary results with the surprise estimated after a Gaussianization process, and for the validity of the calibrated evidence ratio, introduced here for the first time.
        
\section{Discussion}\label{sec:discussion}

In this section we consider three possible origins for the significant tensions we detect between various datasets and the base P15~CMB constraints. First, we discuss the fact that datasets could be affected by systematic effects biasing their constraints; second, we explore the impact on the base P15~CMB constraints of using a flat prior on $\theta_\text{MC}$ instead of on $H_0$ for the curved \LCDM\ model; and finally, we investigate the physical processes underlying the tensions measured in parameter space.

\subsection{Impact of Systematics}\label{sec:systematics}
    
Each of the datasets we consider might be affected by residual systematic uncertainties large enough to lead to tensions with others. As shown elsewhere \citep{seehars1, seehars2}, unresolved systematic uncertainties in the Planck half mission CMB data \citep{planck13} resulted in highly significant tensions with the CMB constraints from WMAP \citep{wmap1, wmap2}, whereas the base P15~CMB constraints are in a far better agreement with WMAP, which holds true also in a series of extended models \citep[see][]{SG}. 
    
To check whether any systematic effect accounted for by the Planck Collaboration might play a role in the 2.7$\sigma$ deviation from flatness, and the 2.3$\sigma$ deviation from $A_\mathrm{L}=1$, in the base P15~CMB data, we show in Fig.~\ref{fig:nuisance} the constraints on the nuisance parameters sampled by the Planck Collaboration with the largest variations between the flat \LCDM\ model (blue lines) and either the curved \LCDM\ (red) or the flat \LCDM$+A_\mathrm{L}$ (green) models. We find no major shifts in the nuisance parameter constraints. Thus, treatment of systematic effects in the base P15~CMB data appears stable under these extensions and not responsible for the tensions reported here. However, this does not exclude the possibility that there are unresolved residual systematics in the P15 data.

Interestingly, the minor shifts induced by the curved \LCDM\ and flat \LCDM$+A_\mathrm{L}$ models are very similar. This hints at a similarity in the way these two models impact the P15~CMB constraints, as discussed in detail in \citet[pg. 29]{planck13} and \citet[]{planck15}.
    
The resulting tensions could also come from the other probes. We discuss the impact of different $H_0$ measurements in Section \ref{sec:cmbvsh0}. For exhaustive discussions of the treatment of systematics in the datasets employed here, we refer the reader to the literature referenced in Section~\ref{sec:data}.

\begin{figure}
\hskip-0.05in
\includegraphics[width=1.0\hsize]{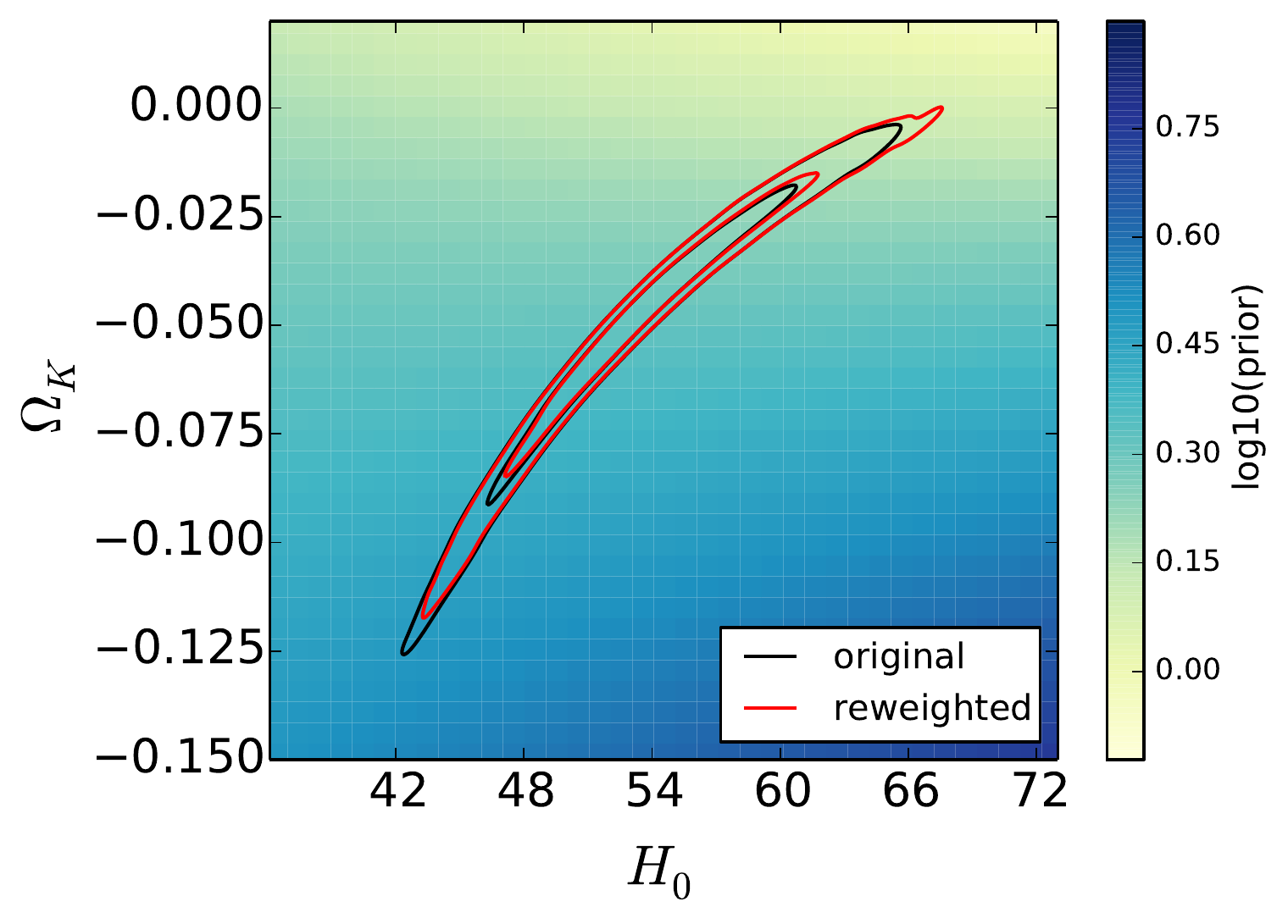}
\vskip-0.1in
\caption{In black, the marginalized contours of the base P15~CMB constraints in curved \LCDM\ over plotting a color-coded background of the $\log_{10}$ of the prior weights, derived from flat priors on both $\Omega_\mathrm{K}$ and $\theta_\text{MC}$. Clearly, the prior puts more weight (up to $10^{0.5}\sim3$) on the low $\rm H_0$, negative $\Omega_\mathrm{K}$ tail of the degeneracy. The red contours are obtained by crudely reweighting the sample (see equation~(\ref{eq:reweight})), to make it correspond to flat priors on $\Omega_\mathrm{K}$ and $H_0$ instead.}
\label{fig:prior}
\end{figure}
%

\subsection{Effect of a Prior Choice}\label{sec:prior}

%
\begin{table}
\caption{Surprise values $S$ and expected fluctuation $\sigma$ for different datasets added to the P15 \texttt{TT\_lowTEB} constraints in curved \LCDM\ after accounting for the reweighing due to the change between using a flat prior on $H_0$ instead of on $\theta_\text{MC}$.}             
\label{table:3}   
\centering                         
\begin{tabular}{c r r r r r r}  
\hline\hline              
& BAO & CMB len. & TEEE & $H_0$ & SNe & Ly$\alpha$ BAO \\     
\hline                       
$S$ & $5.34$ & $4.93$ & $-1.23$ & $6.57$ & $2.49$ & $0.73$ \\      
$\sigma$ & $1.33$ & $1.29$ & $1.08$ & $0.94$ & $0.74$ & $0.45$ \\
$S$/$\sigma$ & $4.02$ & $3.82$ & $-1.14$ & $6.98$ & $3.36$ & $1.62$ \\
\hline                                
\end{tabular}
\end{table}
%

\begin{figure*}
\hskip-0.05in\vskip-0.1in
\includegraphics[width=0.49\hsize]{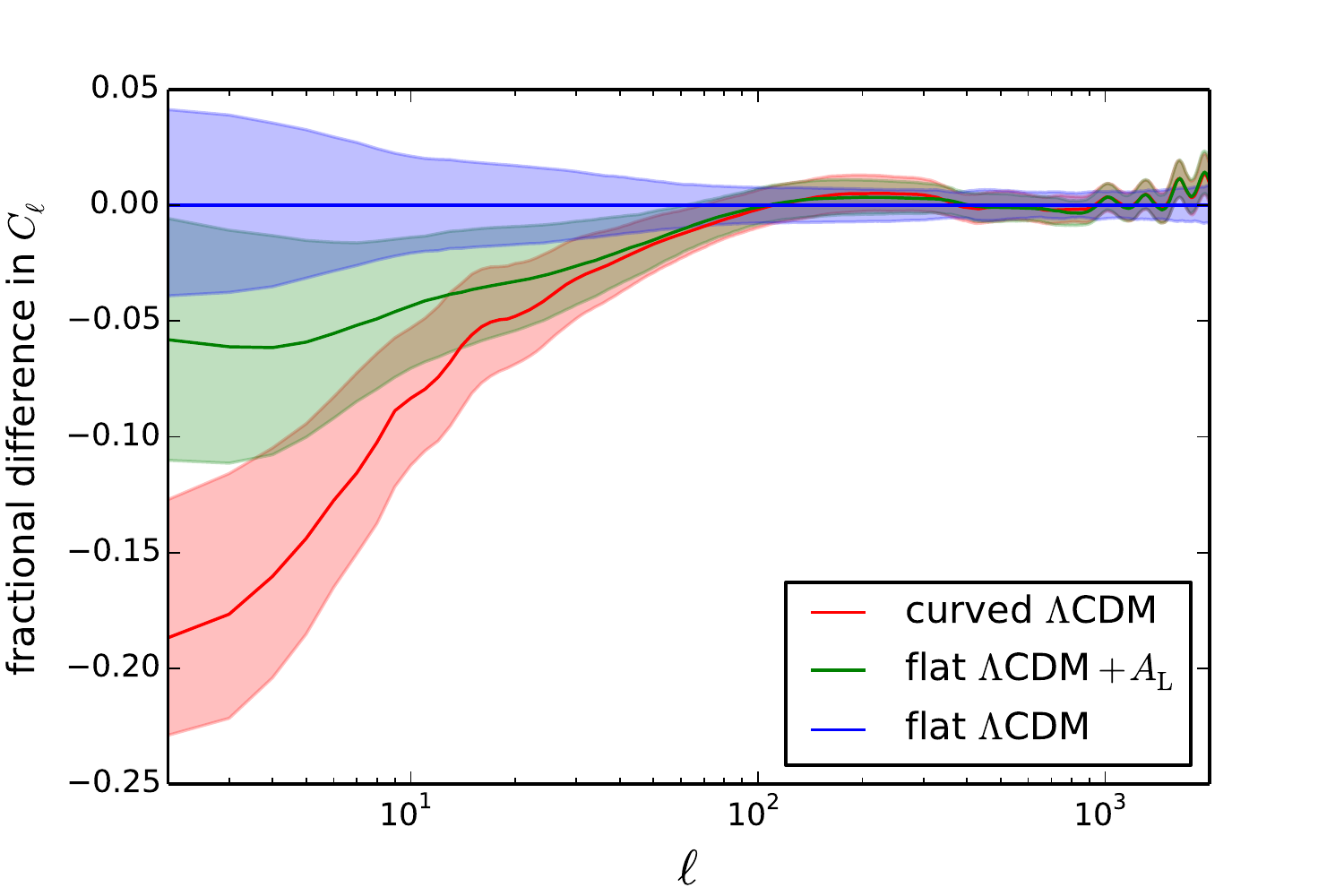}
\includegraphics[width=0.49\hsize]{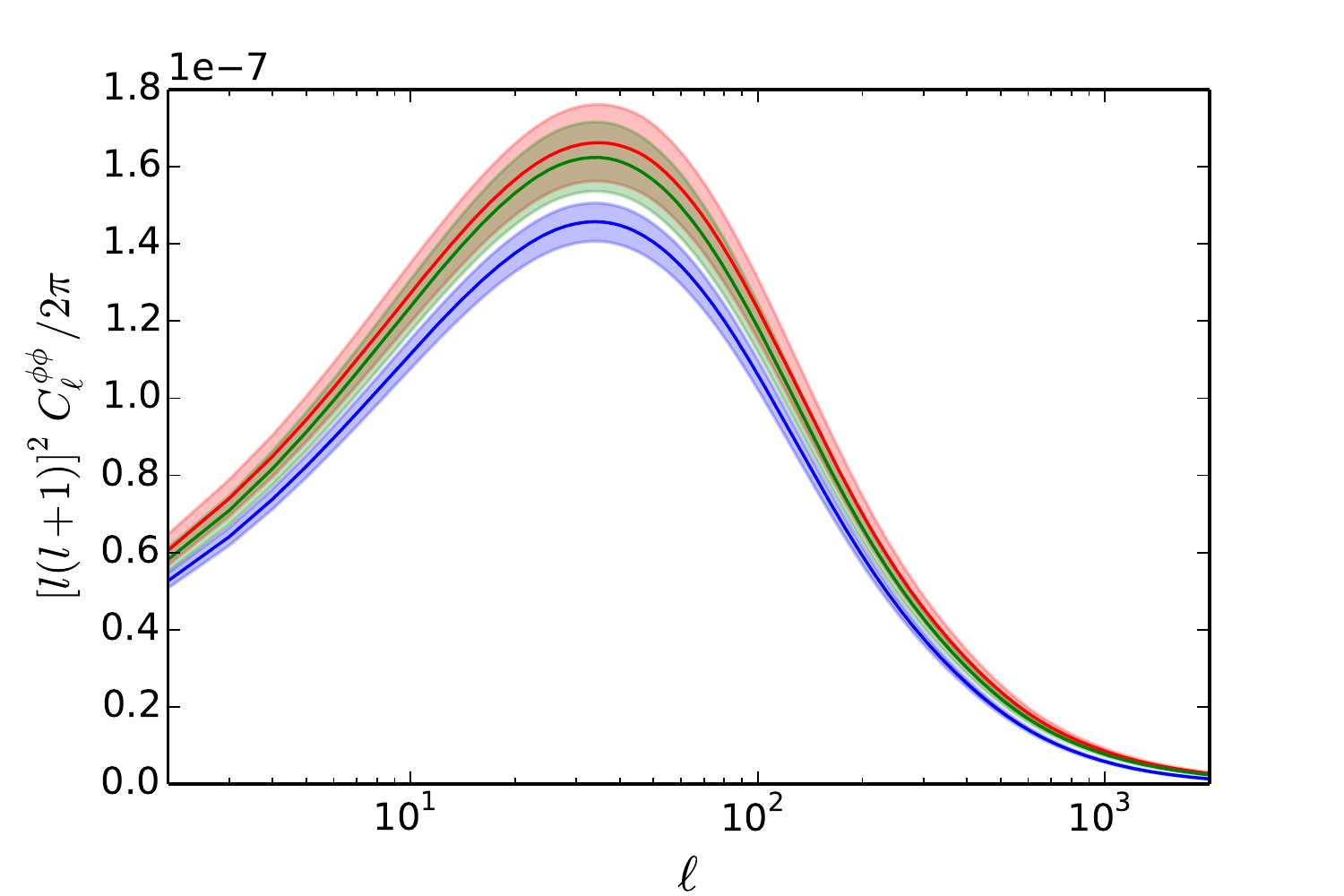}

\vskip-0.1in
\caption{\textit{Left panel:} Fractional differences between the flat \LCDM\ best fit value of the \texttt{TT} power spectrum and those predicted by the constraints obtained in flat \LCDM\ (blue), curved \LCDM\ (red), and flat \LCDM$+A_\mathrm{L}$ (green). For multipole moments $\ell<30$, the P15 temperature anisotropy measurements prefer less power than that predicted by flat \LCDM. This lack of power is stronger in the curved model than in the model with free $A_\mathrm{L}$. \textit{Right panel:} CMB lensing power spectrum predictions from the base CMB constraints obtained in flat \LCDM\ (blue), curved \LCDM\ (red), and flat \LCDM$+A_\mathrm{L}$ (green). Remarkably, the curved and the flat \LCDM$+A_\mathrm{L}$ models predict very similar lensing power spectra, both larger than the prediction from flat \LCDM.}
\label{fig:diff_cls}
\end{figure*}
%

Another effect which could contribute to the preference for non-flat models is the weight assigned to different regions of parameter space by the priors used to sample the base P15~CMB constraints in the curved \LCDM\ model. The Planck Collaboration assumed flat priors on $\Omega_\mathrm{K}$ and $\theta_\text{MC}$. In Fig.~\ref{fig:prior} we show the marginalized contours of the base P15~CMB constraints on the $H_0$, $\Omega_\mathrm{K}$ plane. To crudely estimate the weight of the prior, we fix the other cosmological parameters to their best fit values and compute $\theta_\text{MC}$ on a grid as a function of $H_0$ and $\Omega_\mathrm{K}$ using \texttt{CAMB}. We then numerically compute the flat prior 
\begin{equation}
p(H_0,\,\Omega_\mathrm{K}) = \Big |\frac{\partial (\theta_\text{MC}, \, \Omega_\mathrm{K})}{\partial (H_0,\, \Omega_\mathrm{K})} \Big |\, p(\theta_\text{MC},\,\Omega_\mathrm{K}) \propto \Big | \frac{\partial \theta_\text{MC} }{\partial H_0}\Big |\,,
\label{eq:reweight}
\end{equation}
where $ p(\theta_\text{MC},\,\Omega_\mathrm{K})$ is the prior on $\Omega_\mathrm{K}$ and $\theta_\text{MC}$, which can be assumed $\propto 1$, and $|\partial (\theta_\text{MC}, \, \Omega_\mathrm{K})/\partial (H_0,\, \Omega_\mathrm{K}) |$ stands for the determinant of the Jacobian of the transformation $(\theta_\text{MC}, \, \Omega_\mathrm{K})\mapsto(H_0,\, \Omega_\mathrm{K})$, which can be simplified to $ | \partial \theta_\text{MC} /\partial H_0 |$, the absolute value of the partial derivative of $\theta_\text{MC}$ with respect to $H_0$, evaluated at the relevant position in parameter space.
 
We find that the original priors give more weight to regions away from $\Omega_\mathrm{K}=0$, with up to a factor of $\sim3$ at the low end of the degeneracy, as shown in Fig.~\ref{fig:prior}. We also show there the marginalized contours of the original (in black) and the reweighted (in red) sample obtained from the former using equation~(\ref{eq:reweight}). As an effect of the reweighting, the deviation from flatness is reduced from 2.7$\sigma$ to 2.5$\sigma$. We also calculate numerically the impact of the reweighting on the $\Delta DIC$, finding that it is insignificant and that the clear preference for curved \LCDM\ is maintained. 

In Table~\ref{table:3}, we show the entropy results after reweighting. The significances of the tensions are slightly lower than before reweighting. This comes from the fact that the reweighting pushes the base CMB constraints towards flatness and therefore to better agreement with the other datasets. Nevertheless, as before, with the exception of Ly$\alpha$ BAO, all additional probes maintain more than 3$\sigma$ tension. Thus, we conclude that this change in the prior does not resolve the tensions we find in curved \LCDM\ because it reduces the significances of the tensions and deviations only by $\sim10\%$.  However, it is worth noting that any choice of prior (even flat) in parameter space can indeed introduce unintended preferences for certain regions in this space.

\subsection{CMB and Distance Ladder}\label{sec:cmbvsh0} 

%
\begin{table}
\caption{Surprises $S$ and expected fluctuation $\sigma$ for different $H_0$ measurements when added to the P15 \texttt{TT\_lowTEB} constraints in flat, curved and flat \LCDM $+A_\mathrm{L}$.}
    \vspace{5pt}          
    \label{table:h0s}     
    \centering                     
    \begin{tabular}{c r r r r r r }    
    \hline \hline 
     & \multicolumn{2}{l}{flat \LCDM } &\multicolumn{2}{l}{curved \LCDM } & \multicolumn{2}{l}{flat \LCDM $+A_\mathrm{L}$}\\
 
    \hline         
     & R16 & E14  & R16 & E14  & R16 & E14\\     
    \hline                   
       $S$ & $1.11$ & $0.01$ & $7.36$ & $3.74$  & $0.57$ & $-0.17$  \\  
       $\sigma$ & $0.23$ & $0.08$  & $0.94$ & $0.75$  & $0.37$ & $0.17$ \\
       $S$/$\sigma$ & $4.78$ & $0.09$  & $7.87$ & $4.97$ & $1.52$ & $-1.00$ \\
\hline                             
\end{tabular}
\end{table}
%

The consistency between the Hubble rate inferred from the CMB and distance ladder measurements is a popular and important topic in the recent literature\citep[see for instance][]{h0_tension_1, h0_tension_2, H0concordance, riess16}. Since \citet{planck15} 
adopted $H_0^\text{E14}$ $= 70.6 \pm 3.3\, \text{km}\,\text{s}^{-1}\,\text{Mpc}^{-1}$\citep{efstathiou}, we repeat our analysis with this measurement and obtain the results shown in Table~\ref{table:h0s} for the surprise. E14 agrees better with the base P15~CMB constraints than R16 in all models we considered. For flat \LCDM, E14 is consistent with the CMB constraints, as also found by \citet{planck15}. Compared to previous results from \citet{riess} and \citet{riess14}, the tighter measurements on $H_0$ from R16 are however in significant tension with the P15 CMB constraints even for this simple model.

Interestingly, when we consider the curved \LCDM\ model, all distance ladder measurements show significant tensions with the base P15~CMB constraints. The presence of the tension between the P15~CMB constraints and the distance ladder measurements in the curved \LCDM\ model is thus independent of the specific $H_0$ measurement we choose, although its significance varies (4.97 for E14 and 7.87 for R16). 

To reconcile the constraints on $H_0$ from the CMB and the local distance measures, a variety of mechanisms have been proposed, including an increased $N_\text{eff}$ \citep{2013arXiv1307.0637A, 2016PhRvD..93h3527D}, phantom dark energy \citep{planck13, 2016arXiv160600634D}, or interacting dark energy \citep{2013PhRvD..88b3531S, 2014PhRvD..89j3531C}. It is worth stressing that here we find consistency between the $H_0$ measurements from both E14 and R16 with the base P15 CMB data in the flat \LCDM$+A_\mathrm{L}$ model. Thus, contrary to models with free $N_\text{eff}$ or $w\neq-1$, a model with $A_\mathrm{L}>1$ is not only preferred by the CMB data alone but also provides consistency between these data and all the local distance measures (see additional discussion in Section \ref{sec:model_select}).

\subsection{Physical Effects Involved in the Tensions}  \label{sec:physics}     

To investigate the physical effect causing the deviation from flatness and $A_\mathrm{L}=1$ in the base P15~CMB constraints, we compare the theoretical predictions of the \texttt{TT} spectrum in flat, curved, and flat \LCDM $+A_\mathrm{L}$ models. To do so, we draw random points from the base P15~CMB samples in these models and compute the theoretical expectation of the angular power spectrum of the temperature anisotropies, $C_\ell$, using \texttt{CAMB}. In Fig.~\ref{fig:diff_cls} we show the fractional differences with respect to the best fit values of $C_\ell$ in flat \LCDM. We find that the 1$\sigma$ uncertainty on the flat \LCDM\ prediction (blue region) ranges from $4\%$ at low $\ell$ to less than $1\%$ at high $\ell$, underlining the impressive constraining power of the P15~CMB measurements. For the distribution of the $C_\ell$ in the curved \LCDM\ model (in red), we find that above $\ell\sim 50$ the \texttt{TT} spectra predicted by both models are consistent with each other at the 1$\sigma$ level and within a $2\%$ fractional difference. However, at low $\ell<30$ the curved model is able to predict noticeably less power than the flat model. For the lowest $\ell$, the dipole term, the preferred curved model predicts almost $20\%$ less power than the flat model. As discussed elsewhere \citep[and references therein]{lowl_2, lowl_1}, the lack of power on large scales is one of the anomalies observed in all CMB surveys, P15 included. The 2.7$\sigma$ deviation from flatness seems to be driven by these anomalies and due to the ability of the curved model to predict less power on large scales. Similarly, also the $C_\ell$'s predicted in the model with free $A_\mathrm{L}$ are in excellent agreement with the flat \LCDM\ prediction above $\ell\sim30$. But also in this model, we find a lack of power on large angular scales, although in a less pronounced way than in the curved model. At low redshift, this can be achieved through the Integrated Sachs-Wolfe (ISW) effect \citep[see][]{sachswolfe, sachswolfe2, planck15_de_modgrav}.

However, as discussed by the Planck Collaboration \citep[][pg. 38]{planck15}, the constraints on curvature can also come from an increase of the lensing potential, which directly manifests itself as a deviation of its amplitude $A_\mathrm{L}>1$ \citep[see Section \ref{sec:model_select}, Figure \ref{fig:dics} and pg. 24 in][]{planck15}. To investigate this possibility in further detail, we compute the CMB lensing potential power spectrum $C_\ell^{\phi\phi}$ predicted by the base CMB constraints in flat \LCDM, curved \LCDM\  and flat \LCDM$+A_\mathrm{L}$. The results, together with the 1$\sigma$ uncertainties, are shown in Figure \ref{fig:diff_cls}, where we show the predictions of the CMB lensing power spectra for the flat (blue), curved (red) and flat \LCDM$+A_\mathrm{L}$ (green) models. Remarkably, the curved and the flat \LCDM$+A_\mathrm{L}$ models predict very similar $C_\ell^{\phi\phi}$'s, which are about 2$\sigma$ larger than those predicted by flat \LCDM. From this we conclude that both the deviation from flatness and the deviation from $A_\mathrm{L}$ might be sourced by the same anomaly in the CMB lensing potentials. This might also be supported by the fact that the constraints on the nuisance parameters sampled by P15 are very similar in these two models, as already noted in Section \ref{sec:systematics} (see Fig.~\ref{fig:nuisance}).
    
Although the constraints on the CMB lensing potentials are very similar for the curved and the flat \LCDM$+A_\mathrm{L}$ models and show similar trends in the predicted temperature power spectrum, this is not true for the predicted background evolutions. This manifests itself in our tests of the curved model, where different distance measurements are in significant tension with the CMB. We show that considering the flat \LCDM$+A_\mathrm{L}$ model, the tensions between the base P15~CMB and $H_0$, SNe and BAO are considerably alleviated, both compared to flat and curved \LCDM. Thus, the consistency of the CMB with distance measures in the flat \LCDM$+A_\mathrm{L}$ model seems to suggest that a modification of the CMB lensing potential is preferred to deviations from flatness. However, such modifications to the CMB lensing potential should not only fit the CMB spectra better, they should also be consistent with the CMB lensing measurements, which we find to be in tension with the base CMB data both in the curved and in the flat \LCDM$+A_\mathrm{L}$ models.
    
As shown in \cite{al1} \citep[see also e.g.][]{al2}, $A_\mathrm{L}>1$ is naturally related to theories of modified gravity. Furthermore, the Planck Collaboration \citep{planck15_de_modgrav} reported that the base P15~CMB constraints on some classes of modified gravity models deviate more than 2$\sigma$ from General Relativity. Such models are found to fit the CMB data better than flat \LCDM. It would be interesting to see whether such models can reconcile the CMB lensing measurements with the constraints from the base P15~CMB data.
     
\section{Conclusions}

In this work we first investigate which model is preferred by the CMB temperature and large scale polarization anisotropy measurements of the Planck Collaboration \citep[base P15~CMB;][]{planck15}. Applying the \textit{Deviance Information Criterion} on the posterior samples made publicly available by the Planck Collaboration \citep{planck15}, we find that the base P15~CMB constraints present a strong preference for a \LCDM\ model with free curvature, $\Omega_\mathrm{K}$, over the flat \LCDM\ paradigm. This strong preference comes from the fact that the curved model fits the CMB data at low multipoles ($\ell<30$) better that the flat model, as reported by the Planck Collaboration \citep[][p. 38]{planck15}. We also find that the constraints on $\Omega_\mathrm{K}$ deviate at a 2.7$\sigma$ level from flatness ($\Omega_\mathrm{K}=0$). Furthermore, we find that the base P15~CMB data prefer a model with a CMB lensing potential amplitude $A_\mathrm{L}\neq1$. In this model, the constraints on the additional parameter $A_\mathrm{L}$ are found to deviate from the flat \LCDM\  expectation ($A_\mathrm{L}=1$) by 2.3$\sigma$. If this result is not due to residual systematics in the data, our model selection analysis (see Section~\ref{sec:model_select}) indicates that it represents a challenge to the standard flat \LCDM\ model.
        
To investigate whether there is concordance between different measurements in these models, we consider the addition of external datasets to the base P15~CMB constraints. We utilize the joint constraints published by the Planck Collaboration \citep{planck15} from measurements of the base P15~CMB together with CMB lensing, CMB small scale polarization, BAO, SNe, distance ladder or Ly$\alpha$ forest BAO.  To analyze these datasets, we simultaneously \textit{Gaussianize} the constraints from the base P15~CMB data and the combined datasets, and obtain an analytic approximation to their likelihood that enables the calculation of the entropy based measure \textit{surprise} \citep{seehars1, seehars2, SG} and a \textit{calibrated} evidence ratio, as well as a more efficient evaluation of the likelihood.

In the flat \LCDM\ model, we find that all external datasets agree with the base P15~CMB, except for the distance ladder measurement performed by \citet{riess16}, which we find to be in 4.8$\sigma$ tension. In the curved \LCDM\ model, which is clearly preferred by the base P15~CMB data, we find significant tensions between the CMB and distance ladder (7.9$\sigma$), BAO (4.6$\sigma$), CMB lensing (4.2$\sigma$) and SNe (3.8$\sigma$) measurements. The curved model is thus unable to describe these observations adequately. Given these high levels of tension, these datasets should not currently be added to the base P15~CMB constraints in the curved model until these inconsistencies can be resolved. Considering instead a model with a free CMB lensing potential amplitude $A_\mathrm{L}$, the base P15~CMB constraints are consistent with the different distance measures, even resolving the tensions between the CMB and distance ladder measurements. However, in this model the CMB lensing measurements are still in about $4\sigma$ tension with the base P15~CMB data. 

Using a simple example, we also show the importance of accurately calibrating the evidence ratio to have an unbiased assessment of the consistency between two datasets. To validate our primary measure of tension, we introduce the \textit{calibrated evidence ratio} and calculate its expected fluctuation. Applying this measure to some of the datasets gives us significances of the tensions that are in good agreement with those from the surprise. 
    
We also discuss the possible effects driving the deviation from flatness in the base P15~CMB constraints and therefore the tensions of these data with different external datasets. Our examination uncovers no evidence that these are due to systematics currently accounted for in the CMB analysis; however, we cannot exclude that these are due to unresolved, residual systematics. Also, the choice of using a flat prior on $\theta_\text{MC}$ instead of $H_0$ for the CMB analysis introduces only a $10\%$ bias on the reported significances of the deviations and tensions, and is thus insufficient to explain them. 
    
We also compute the \texttt{TT} spectra predicted by the base CMB constraints in the flat model and in the preferred models with free curvature and lensing amplitude. When comparing them to flat \LCDM, we find a lack of power on large scales of almost $20\%$ for the curved, and $5\%$ for the $+A_\mathrm{L}$ model, respectively. Large scale lack of power has been consistently found in all CMB all-sky surveys, and might source the deviation we find here. This anomaly partially manifests itself as an increment of the CMB lensing potential. Remarkably, both the curved and the flat \LCDM$+A_\mathrm{L}$ models predict larger CMB lensing potentials than the flat \LCDM\ model. However, the curved model increases the lensing potentials at the cost of altering the cosmological background in a way that is incompatible with external distance measurements. On the other hand, a model that impacts the CMB lensing potentials without significantly changing the background expansion would allow consistency between the base P15~CMB data and external distance measurements. Such an alternative model should also be able to reconcile the direct CMB lensing measurements with the constraints coming from the temperature anisotropy power spectrum, which is not the case with the flat \LCDM$+A_\mathrm{L}$ model, as we have shown here. The important ongoing efforts in measuring the cosmic large scale structure in large survey projects such as, for example, DES\footnote{\url{http://www.darkenergysurvey.org}} \citep{DES05}, eROSITA\footnote{\url{http://www.mpe.mpg.de/eROSITA}} \citep{merloni12}, EUCLID\footnote{\url{http://sci.esa.int/euclid/}} \citep{laureijs11} and LSST \footnote{\url{http://www.lsst.org}} \citep{LSST09} will provide us with additional consistency checks among datasets while yielding tighter constraints that enable further systematic tests of alternative models to flat \LCDM.

\section*{Acknowledgements}
SG thanks Alexander Refregier, Adam Amara and Sebastian Seehars for fruitful discussions on various aspects of this work. We also thank the anonymous referee for useful comments.  We acknowledge the support by the DFG Cluster of Excellence ``Origin and Structure of the Universe'', the Transregio program TR33 ``The Dark Universe'' and the Ludwig-Maximilians-Universit\"at. DR is currently supported by a NASA Postdoctoral Program Senior Fellowship at the NASA Ames Research Center, administered by the Universities Space Research Association under contract with NASA. We acknowledge use of the Planck Legacy Archive. Planck (\url{http://www.esa.int/Planck}) is an ESA science mission with instruments and contributions directly funded by ESA Member States, NASA, and Canada.


\bibliographystyle{mnras}
\bibliography{my_bib}


\appendix

\section{Gaussianization Procedure}\label{sec:gaussianization}

\begin{figure*}
\hskip-0.10in
 \includegraphics[width=\textwidth]{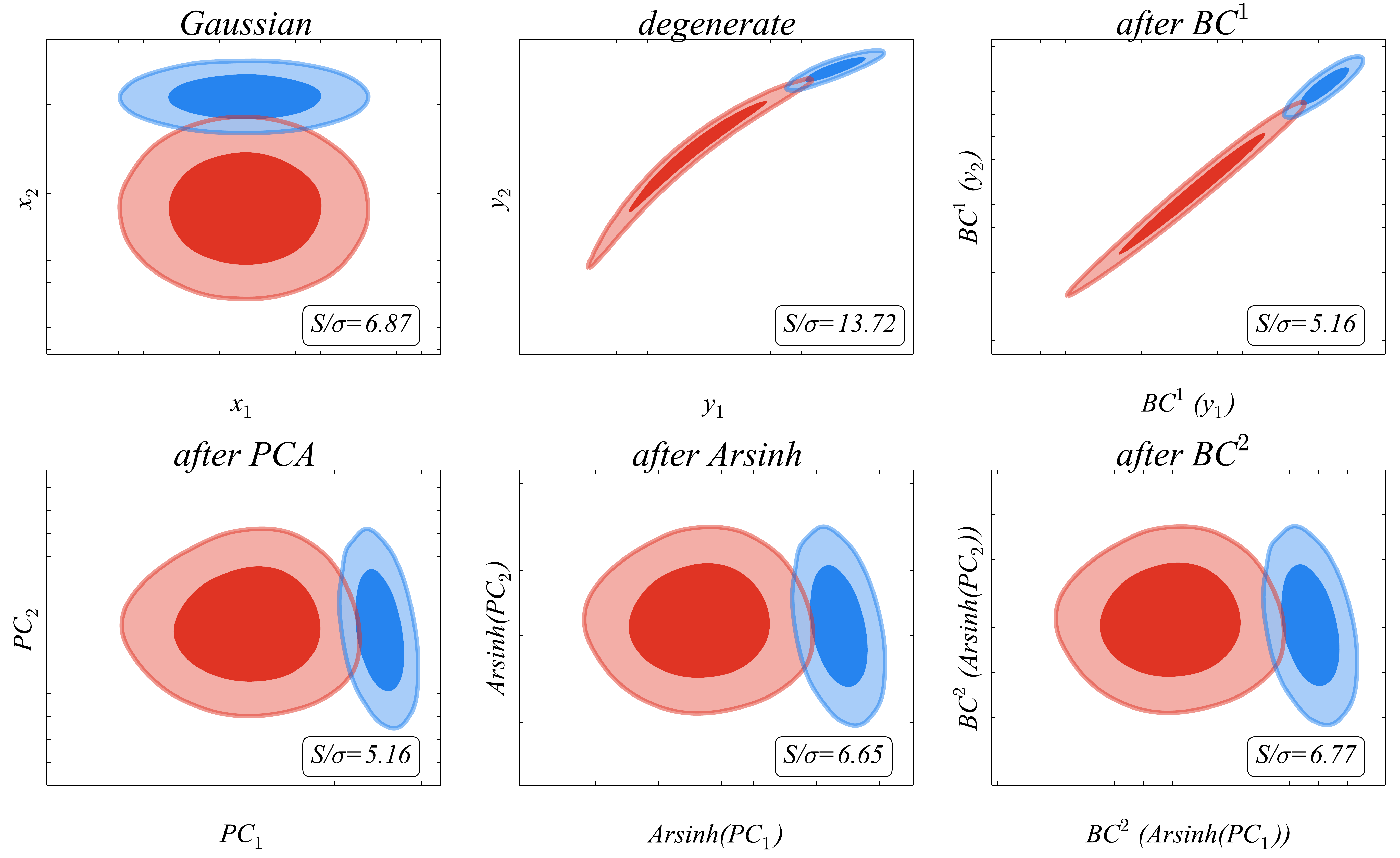}
 \vskip-0.1in
\caption{Test case, illustrating the Gaussianization procedure. The upper left panel shows the original, Gaussian samples, for which the significance of the tension $S/\sigma$ can be computed analytically. By applying a non linear mapping, these samples can be transformed into the degenerate constraints shown in the upper middle panel. The subsequent panels (upper right, lower left, lower middle and lower right) show the samples after applying the Gaussianizing transformations presented in Table~\ref{table:trans} (except for the first). The final significance is very close to the correct, initial value, indicating that the samples are well approximated by Gaussian distributions.}
\label{fig:test_case}
\end{figure*}

%

    Following \citet{schuhmann}, we compute a suite of optimized transformations to Gaussianize a distribution. We first apply a linear transformation $\mtrx{M}$, mainly to decorrelate strongly degenerate parameters. Thereafter, we apply a BoxCox transformation to each dimension individually. A BoxCox transformation is defined by
    \begin{equation}
    \text{BC}_{(a, \lambda)}(x) = 
        \begin{cases}
             \frac{1}{\lambda} (x+a)^\lambda -1 & \text{ if } \lambda \neq 0 \\
             \log(x+a)  & \text{ if } \lambda = 0 \,.
         \end{cases}
    \end{equation}
The optimal transformation parameters are found by maximizing the probability that the transformed sample is Gaussian. This transformation is only defined for $x+a>0$, so given an optimal $a$, the transformation is not defined for all $x$. However, we always choose $a > \text{max}(-x_i)$ for a sample $x_i$ such that the transformation is defined for every point of the sample, but not in every point of parameter space. For a sufficiently large sample, however, we can assume that the value of the probability density distribution is arbitrarily close to zero in regions without sample points. 
    
After the first BoxCox transformation, we apply a principal component analysis (PCA), re-centering the sample by its mean $\pmb{\mu}_\text{PCA}$ and applying a linear transformation $\mtrx{L}^{-1}$ such that after the transformation the sample is standardized. The linear transformation can be obtained from a Cholesky decomposition of the covariance matrix $\mtrx{C}_\text{PCA} = \mtrx{L}\, \mtrx{L}^T$.

After the PCA, we perform another family of transformations. Inspired by \citet{schuhmann}, we apply an Arsinh transformation defined by 
    \begin{equation}
    \text{Arsinh}_{(b, t)}(x) = 
        \begin{cases}
             \frac{1}{t} \sinh(t(x-b)) & \text{ if } t > 0 \\
             x-b  & \text{ if } t = 0 \\
             \frac{1}{t} \text{arsinh}(t(x-b)) & \text{ if } t > 0 \,.
         \end{cases}
    \end{equation}
The transformation is applied again to each dimension individually. The optimal transformation parameters are determined by maximizing the probability that the transformed sample is Gaussian, as done by \citet{schuhmann}. The Arsinh transformation is helpful, because it can transform away some excess kurtosis.
    
    As the last transformation step, we apply again a BoxCox transformation. At this point, for our cases the samples we consider are well approximated by a Gaussian. Thus, we estimate the final mean $\pmb{\mu}_\text{final}$ and the final covariance $\mtrx{C}_\text{final}$. Table~\ref{table:trans} summarizes the transformations and the transformation parameters necessary in every point.

\begin{table}
    \caption{Summary of the transformations (trans.) employed to Gaussianize a generic sample. We also specify the transformation parameters (params.) for each transformation. The index $i$ runs from $1$ to $n_\text{dim}$, which is the number of dimensions.}  
    \vspace{5pt}        
    \label{table:trans}     
    \centering                    
    \begin{tabular}{c c c }   
    \hline\hline              
     & trans. & params. \\   
    \hline                   
    1st & linear & $\mtrx{M}$ \\
    2nd & BoxCox & ($a^{(1)}_i$, $\lambda^{(1)}_i$)  \\
    3rd & PCA & $\pmb{\mu}_\text{PCA}$, $\mtrx{C}_\text{PCA}$\\
    4th & Arsinh & $(b_i, t_i)$ \\
    5th & BoxCox & ($a^{(2)}_i$, $\lambda^{(2)}_i$)  \\
    \hline                             
    \end{tabular}
\end{table}
%

   The Gaussianization procedure gives an analytic approximation to the distribution from which the original sample has been drawn. Any point in cosmological parameter space $\pmb{\theta}$ needs to be transformed by the transformations shown in Table~\ref{table:trans}, yielding $\pmb{\psi} = \text{trans}(\pmb{\theta})$. Then its likelihood can be approximated by using the expression derived by \citet{sellentin}, accounting for the scatter introduced by estimating the covariance of the sample. Using this method, we obtain analytic approximations for the P15~CMB likelihood for the models we consider. We make various of these products publicly available on \url{https://bitbucket.org/grandiss45/gaussianization/}.

Optimising the above given suite of transformations to optimally Gaussianize two samples allows one to jointly Gaussianize two distributions. A joint Gaussianization is theoretically not possible in general, but for prior and posterior distributions, a joint Gaussianization is feasible, because the posterior is generally better behaved than the prior. This allows us to estimate the surprise and its significance analytically using equation~(\ref{eq:S}).

\subsection{Test Case}
To provide an example of the Gaussianization procedure and compute the accuracy with which we can estimate the significance of tensions after Gaussianization, we construct the following test case. We start from two Gaussian distributions, shown in the upper left panel of Figure~\ref{fig:test_case}, for which we can compute the significance of the tension analytically. Applying a non linear transformation different from those in Table~\ref{table:trans}, we transform these samples to the degenerate constraints shown in the upper middle panel of the figure. This transformation is defined as follows. Given the two components of the original Gaussian samples, $x_{1,2}$, we transform them into $y_1 = x_1^\beta (x_2^\alpha+C)$ and $y_2 = x_1^\beta/(x_2^\alpha+C)$, where we choose $\alpha=1.1$, $\beta=0.4$, and $C=4$. This transformation cannot be obtained analytically from the Gaussianizing transformations and therefore defines an interesting test case for the Gaussianization procedure. Furthermore, this simple case has a clear similarity to the P15 CMB constraints in the curved model.

As expected, the significance of the tension estimated from these samples is highly inaccurate, because the estimation of the surprise and its variance assumes that the distributions are Gaussian. In fact, all entropy derived quantities are invariant under arbitrary invertible parameter transformations, but our estimation assumes that the samples are Gaussian. Besides statistical uncertainty due to the finiteness of the samples, any systematic uncertainty in the estimation of the surprise derives from the fact that the underlying samples are not accurately described by Gaussian distributions.

By construction, after every Gaussianizing transformation, the samples are more accurately described by Gaussians and consequently the estimation of the significance is more accurate (see Figure~\ref{fig:test_case}). After applying all the transformations we choose to perform (i.e. those in Table~\ref{table:trans} with the exception of the first), the fractional accuracy on the significance is $1.5\%$. Note that the final Gaussianized parameter space need not be equal to that of the initial Gaussian distributions.

\subsection{Accuracy of the Gaussianization}

To estimate the accuracy of the significances of the underlying tensions of the cosmological constraints analyzed in this work we propose the following scheme.

As described above, the accuracy of the significance depends solely on the degree to which the samples are well described by Gaussians. A natural measure of how well a distribution $\hat p$ describes a sample $\mathcal{X}=\{ x_i\}$, with $i=1, ..., N$, where $N$ is the length of the sample, is given by the \textit{logarithmic score}

\begin{equation}
H_\mathcal{X}^{\hat p} = \frac{1}{N}\sum_{i=1}^N \ln \hat p (x_i).
\end{equation}
If the sample points are drawn independently, the logarithmic score can be interpreted as the average log-likelihood that the sample $\mathcal{X}$ has been drawn from $\hat p$.\footnote{Given the likelihood $L(x_i|\hat p)=\hat p(x_i)$ that the sample points $x_i$ are drawn independently from $\hat p$, the likelihood that the sample $\mathcal{X}$ is drawn from $\hat p$ is $L(\mathcal{X}|\hat p)=\prod_i \hat p(x_i)$. Thus, $\ln L(\mathcal{X}|\hat p) = \sum_i \ln \hat p (x_i) = N\, H_\mathcal{X}^{\hat p}$.} Consequently, a higher logarithmic score indicates a better fit.

For the case of the P15 CMB constraints in curved \LCDM, labelled hereinafter $pr$ for prior, and the joint constraints of P15 CMB and CMB lensing data, labelled $po$ for posterior, we define $\hat q$ and $\hat p$, the approximations to the prior and posterior in the space of parameters after the Gaussianitation process, as Gaussian likelihoods with means and covariances estimated from the transformed samples. We then draw 2000 samples $\mathcal{X}^{pr}$ and $\mathcal{X}^{po}$ from $\hat q$ and $\hat p$, respectively, and compute the logarithmic scores $H^{\hat q}$ and $H^{\hat p}$. We also evaluate the logarithmic scores of the original samples $H_{pr}^{\hat q}$ and $H_{po}^{\hat p}$. We find that
\begin{equation}
H_{pr}^{\hat q}=-3.000 \text{ and } \langle H^{\hat q} \rangle = -2.998 \pm 0.007\,, 
\end{equation}
for the prior, and
\begin{equation}
H_{po}^{\hat p}=-3.000 \text{ and } \langle H^{\hat p} \rangle = -2.994 \pm 0.006\,, 
\end{equation}
for the posterior. We note that for both, the prior and the posterior, the Gaussian samples $\mathcal{X}^{pr, po}$ on average fit better than the original samples $pr$ and $po$. The logarithmic scores, however, are consistent with those of these Gaussian samples within the statistical uncertainties of the sampling process. It is thus safe to assume that the original samples are fitted by $\hat q$ and $\hat p$ to an accuracy consistent with the statistical noise of samples of their size.

To evaluate how large the impact of this statistical sampling noise is on the errors of estimating the significance, for each of the 2000 cases we estimate the significance $S/\sigma$ of the tension between the sample drawn from $\hat q$ and the sample drawn from $\hat p$. We find that the average $\left<S/\sigma\right> = 4.23 \pm 0.03$. For the case of P15 CMB versus P15 CMB plus CMB lensing in curved \LCDM\ we have from Table~\ref{table:surprise} that $S/\sigma=4.18$. This implies an average absolute error of $0.05$, which corresponds to a fractional error of $1.1\%$. Note that this should be interpreted as a systematic error. For the other data combinations and models, since this case is in no way special, we expect similar results.

In summary, the Gaussianization process is successful within the statistical uncertainties of the samples, and introduces systematic errors of the order of only $1\%$, thus allowing a robust inference of the significance of an underlying tension. 
  
\section{Evidence Ratios} \label{sec:app_b}

    It is common practice in cosmology to use $\ln R$, as derived from equation~(\ref{eq:R}), to assess the agreement between two datasets $D_1$ and $D_2$, and $\ln R = 0$ is used as a reference point for such assessments. However, $\ln R$ depends on $D_1$ and $D_2$, which themselves are random variables, making also $\ln R$ a random variable. Consequently, in this section we follow the reasoning of \citet{seehars2} and for a class of likelihood models we propose a statistically well motivated reference point. We also analyze the statistical scatter of the measure $\ln R$.
    
    \subsection{Statistics in One Dimension} \label{sec:evidence}
    
    We first consider a simple one dimensional model with a flat prior $p(\theta)$ and likelihood 
    \begin{equation}\label{eq:example_likeli}
    L(D_i|\,\theta) = \frac{1}{\sqrt{2 \pi s_i^2}} \exp \Bigg (-\frac{1}{2}\,\left(\frac{\theta-D_i}{s_i}\right)^2\Bigg ) \text{ for } i=1,\,2\,, 
    \end{equation}
    where $s_i$ are the uncertainties of the datasets. These likelihoods are normalized in a way that $E(D_i)=1$. This model accurately describes the constraints on $H_0$ from the CMB and distance ladder measurements used here both in flat and curved \LCDM, and the constraints of SNe and CMB on $\Omega_M$ in flat \LCDM.
    
    In this setting, the joint distribution of the parameter $\theta$ and the datasets $D_1$, $D_2$ is given by
    \begin{equation} \label{eq:pthetad1d2}
    \begin{split}
    p(\theta,\, D_1, \, D_2) = \frac{1}{2\pi\sqrt{ s_1^2 s_2^2}} & \exp \Bigg (-\frac{1}{2}\,\frac{(D_1-D_2)^2}{s_1^2+s_2^2}\Bigg ) \\
    &\exp \Bigg (-\frac{1}{2}\,\frac{(s_1^2+s_2^2)\, ( \theta - \mu )^2}{s_1^2 s_2^2}\Bigg )\,, 
    \end{split}
    \end{equation}
    with $\mu=(s_2^2\,d_1 + s_1^2\,d_2)/(s_1^2+s_2^2)$. Marginalising the expression~(\ref{eq:pthetad1d2}) over the parameter $\theta$ with the flat prior gives the joint evidence of $D_1$, $D_2$ in the form
    \begin{equation} \label{eq:ed1d2}
    E(D_1, D_2) = \frac{1}{ \sqrt{2\pi (s_1^2 +s_2^2)}} \exp \Bigg (-\frac{1}{2}\,\frac{(D_1-D_2)^2}{s_1^2+s_2^2}\Bigg )\,, 
    \end{equation}
    which illustratively is a Gaussian distribution of the difference between the datasets $\Delta D = D_1-D_2$, with variance given by the sum of the variances of the single datasets. Note also that dividing equation~(\ref{eq:pthetad1d2}) by equation~(\ref{eq:ed1d2}) gives the posterior distribution  $p(\theta|\,D_1, D_2)$, which consistently has expected value $\mathbb{E}[\theta|\,D_1, D_2] = \mu = (s_2^2\,d_1 + s_1^2\,d_2)/(s_1^2+s_2^2)$ and variance $\text{Var}[\theta|\,D_1, D_2] = s_2^2 s_1^2/(s_1^2+s_2^2)$.
    
    Using equation~(\ref{eq:ed1d2}) and equation~(\ref{eq:R}) we can compute $\ln R$ analytically 
    \begin{equation}\label{eq:lnr}
    \ln R = -\frac{1}{2}\frac{\Delta D^2}{s_1^2+s_2^2} - \frac{1}{2}\ln (s_1^2+s_2^2) - \frac{1}{2} \ln (2 \pi)\,.
    \end{equation}
    From this expression, it becomes clear that perfectly agreeing datasets ($\Delta D = 0$) will have $\ln R < 0$ to a degree depending mainly on the measurement uncertainties. For example, one could obtain $\ln R = -6$, when comparing the two measurements $D_1 = D_2 = 0 \pm 114$. Using Jeffreys' scale for the natural logarithm, we would describe these results as the datasets being in `strong disagreement', but in fact the data could not agree better! This example should clarify the importance of calibrating $\ln R$ correctly. In the same spirit as that used to calibrate the relative entropy, we propose $\langle \ln R \rangle_{D_1, D_2}$, the expected evidence ratio, as the reference point from which to assess the agreement between two datasets. In our simple model this quantity can be computed analytically as follows  
    \begin{equation}\label{eq:lnr_exp}
    \begin{split}
    \langle \ln R \rangle_{D_1, D_2} &= \int \text{d} D_1\,\text{d}D_2\, E(D_1, D_2)\, \ln R =\\
    &    = - \frac{1}{2}\ln (s_1^2+s_2^2) - \frac{1}{2} \ln (2 \pi) - \frac{1}{2}\,.
    \end{split}
    \end{equation}
Combining equations~(\ref{eq:lnr}) and (\ref{eq:lnr_exp}), we find that the calibrated evidence ratio is given by
    \begin{equation}\label{eq:cal_lnr}
    \begin{split}
     \ln R - \langle \ln R \rangle_{D_1, D_2} &= -\frac{1}{2}\frac{\Delta D^2}{s_1^2+s_2^2} + \frac{1}{2}\,, 
    \end{split}
    \end{equation}
    which effectively cancels the second term of equation~(\ref{eq:lnr}), which depends on the dataset uncertainties. Applying this calibrated evidence ratio to the previous example we find $\ln R - \langle \ln R \rangle_{D_1, D_2} = 1/2$, so a better agreement than statistically expected.   
      
    Equation~(\ref{eq:cal_lnr}) also allows a direct comparison of the calibrated evidence ratio and the surprise, because both are normalized and have scatter around 0. There is, however, a subtle difference in the way the surprise and the calibrated evidence ratio spot tensions between two datasets $D_1$, $D_2$. The calibrated evidence ratio is a symmetric measure of the consistency between the two datasets in data space. It considers directly the  square difference between the datasets compared to the sum of their variances. The surprise is not symmetric and acts in parameter space, as can be seen in equation~(\ref{eq:S}). Instead, it considers the agreement between $p(\pmb{\theta}|\,D_1, M)$ and $p(\pmb{\theta}|\,D_2,D_1, M)$, and assesses how probable the difference between $p(\pmb{\theta}|\,D_1, M)$ and $p(\pmb{\theta}|\,D_2,D_1, M)$ is. It goes after the question: given $D_1$, how probable is it that $D_2$ shifts the mean values of $p(\pmb{\theta}|\,D_1, M)$ to the mean value of $p(\pmb{\theta}|\,D_2,D_1, M)$? Consequently, it is suited to test whether $D_2$ should be added to the constraints of $D_1$, which is in general different from the question of adding $D_1$ to $D_2$.
    
    As with the surprise, we can also derive an expected fluctuation of the calibrated evidence ratio $\sigma(\ln R)$
    \begin{equation}\label{eq:sigma_lnr}
    \begin{split}
    \sigma^2(\ln R) &= \Big \langle \big (\ln R - \langle \ln R \rangle \big)^2  \Big \rangle_{D_1, D_2} = \\
    &= \Bigg \langle \Bigg (-\frac{1}{2}\frac{\Delta D^2}{s_1^2+s_2^2} + \frac{1}{2} \Bigg )^2 \Bigg \rangle_{D_1, D_2} = \frac{1}{2}.
    \end{split}
    \end{equation}
     Thus, in the previous example, the calibrated evidence ratio has a significance 0.7$\sigma$. Calibrating and calculating the scatter of the $\ln R$ for more general likelihoods and priors, however, might require costly numerical computations. For this reason, we prefer the surprise as a measure of tension in the current analysis.
     
    \subsection{Estimation for Gaussian Likelihoods}\label{sec:est_ev}

For a Gaussian likelihood such as that in equation~(\ref{eq:example_likeli}), which approximates the distance ladder measurements of $H_0$ in flat and curved \LCDM\ and the SNe constraints on $\Omega_M$ in flat \LCDM, we have $E(D_1)=1$. If we want to compute the evidence ratio between these and the base P15~CMB dataset, $D_2$, we can use the fact that
    \begin{equation} \label{eq:est_ev_1}
    \begin{split}
    R &= \frac{E(D_1, D_2)}{E(D_1)\, E(D_2)} = E(D_1|\,D_2) = \int \text{d}\theta \, L(D_1|\,\theta)\,p(\theta|\,D_2)\,,
    \end{split}
    \end{equation}
    where $p(\theta|\,D_2)$ is the posterior derived from $D_2$  \citep[for a proof see][]{seehars2}. Given a sample of $p(\theta|\,D_2)$, and an analytic expression for  $L(D_1|\,\theta)$, equation~(\ref{eq:est_ev_1}) can be estimated with Monte Carlo Integration. 

    \subsection{Calibrating Evidence Ratios in N Dimensions}
    For completeness, we give here the $n$-dimensional generalization of the equations given in Appendix~\ref{sec:evidence}. Assume a linear likelihood model for the datasets $D_i$, $i=1,2$ given by
    \begin{equation} \label{eq:b9}
    L(D_i|\,\pmb{\theta}) = \frac{1}{\sqrt{(2\pi)^n\,\det\Sigma_i}}\exp \Big ( -\frac{1}{2} (\pmb{\theta}- \pmb{\mu}_i)^T \Sigma_i^{-1}  (\pmb{\theta}- \pmb{\mu}_i) \Big )\,,
    \end{equation}
    where $\pmb{\theta}$ is the $n$-dimensional model parameter vector, $\pmb{\mu}_i$ a $n$-dimensional vector depending linearly on the dataset $D_i$, and $\Sigma_i$ are symmetric $n \times n$ matrices, independent of the dataset $D_i$ and the model $\pmb{\theta}$.
    
    Integrating equation~(\ref{eq:b9}) over a flat prior $p(\pmb{\theta})=1$, we find the evidence $E(D_i) = 1$. Applying Bayes Theorem, we obtain the posterior distributions $p(\pmb{\theta}|\,D_i) =  L(D_i|\,\pmb{\theta})$. Thus, $\pmb{\mu}_i
    $ is the mean of the posterior $p(\pmb{\theta}|\,D_i)$, and $\Sigma_i$ its covariance. 
    
    Performing the same calculations as in the one dimensional case, we find the joint evidence
    \begin{equation}
    E(D_1, D_2) = \frac{1}{\sqrt{(2\pi)^n \det(\Sigma_1+\Sigma_2)}}\exp \Big ( -\frac{1}{2} \pmb{\Delta \mu}^T (\Sigma_1+\Sigma_2)^{-1}  \pmb{\Delta \mu} \Big )\,.
    \end{equation}
    where $\pmb{\Delta \mu} = \pmb{\mu}_1- \pmb{\mu}_2$ is the difference in means of the posterior distributions $p(\pmb{\theta}|\,D_{1,2})$ . This form is a manifest generalization of equation~(\ref{eq:ed1d2}). In the same way as described above, we can derive the evidence ratio
    \begin{equation}
    \ln R = -\frac{1}{2}\pmb{\Delta \mu}^T (\Sigma_1+\Sigma_2)^{-1} \pmb{\Delta \mu} - \frac{n}{2}\ln 2\pi -\frac{1}{2}\ln \det (\Sigma_1+\Sigma_2)\,.
    \end{equation}
    We can thus confirm that also the $n$-dimensional evidence ratio scatters around a term that depends on the covariance. To find the correct zero point, we need to calibrate it by subtracting its expected value. This gives the $n$-dimensional calibrated evidence ratio 
    \begin{equation}
    \ln R - \langle \ln R \rangle = -\frac{1}{2}\pmb{\Delta \mu}^T (\Sigma_1+\Sigma_2)^{-1} \pmb{\Delta \mu} + \frac{n}{2}\,, 
    \end{equation}
    with a variance Var$[\ln R] = n/2$.
    
    Since the evidence is invariant under parameter transformations, these quantities could be easily estimated after a joint Gaussianization of the two independent posteriors $p(\pmb{\theta}|\,D_1)$ and $p(\pmb{\theta}|\,D_2)$. Here we did not use this method because we had a simpler access to the joint posteriors $p(\pmb{\theta}|\,D_1, D_2)$, which are in general better behaved and thus easier to Gaussianize.

\end{document}